\documentclass[fleqn,usenatbib]{mnras}

\usepackage{newtxtext,newtxmath}

\usepackage[T1]{fontenc}

\DeclareRobustCommand{\VAN}[3]{#2}
\let\VANthebibliography\thebibliography
\def\thebibliography{\DeclareRobustCommand{\VAN}[3]{##3}\VANthebibliography}

\usepackage{graphicx}	
\usepackage{amsmath}	

\usepackage{epstopdf}
\usepackage{verbatim}
\usepackage{natbib}
\usepackage{hyperref}
\usepackage{float}
\usepackage{color}
\usepackage{footnote}
\makeatletter
\newlength{\abovecaptionskip}%
\makeatother
\usepackage{threeparttable}%

\title[Testing IC/CMB in Resolved Quasar X-ray Jets]{\textit{A Multi-Wavelength Study of Multiple Spectral Component Jets in AGN: Testing the IC/CMB Model for the Large-Scale-Jet X-ray Emission}}

\author[P. Breiding et al.]{
Peter Breiding,$^{1}$\thanks{E-mail: pbreiding@gmail.com}
Eileen T. Meyer,$^{2}$
Markos Georganopoulos,$^{2,3}$
Karthik Reddy,$^{2}$
Kassidy E. Kollmann,$^{2,4}$
\newauthor and Agniva Roychowdhury$^{2}$
\\
$^{1}$Department of Physics and Astronomy, Bloomberg Center, Johns Hopkins University, Baltimore, MD 21218, USA\\
$^{2}$Department of Physics, University of Maryland Baltimore County, Baltimore, MD 21250, USA\\
$^{3}$NASA Goddard Space Flight Center, Code 663, Greenbelt, MD 20771, USA\\
$^{4}$Department of Physics, Princeton University, Princeton, NJ 08544, USA
}

\date{Accepted XXX. Received YYY; in original form ZZZ}

\pubyear{2022}

\begin{document}
\label{firstpage}
\pagerange{\pageref{firstpage}--\pageref{lastpage}}
\maketitle

\begin{abstract}
Over $\sim$150 resolved, kpc-scale X-ray jets hosted by active galactic nuclei have been discovered with the \textit{Chandra} X-ray Observatory.  A significant fraction of these jets have an X-ray spectrum either too high in flux or too hard to be consistent with the high-energy extension of the radio-to-optical synchrotron spectrum, a subtype we identify as Multiple Spectral Component (MSC) X-ray jets.  A leading hypothesis for the origin of the X-rays is the inverse-Compton scattering of the cosmic microwave background by the same electron population producing the radio-to-optical synchrotron spectrum (known as the IC/CMB model).  In this work, we test the IC/CMB model in 45 extragalactic X-ray jets using observations from the \textit{Fermi} Large Area Telescope to look for the expected high level of gamma-ray emission, utilizing observations from the Atacama Large Millimeter/submillimeter Array (ALMA) and the \textit{Hubble Space Telescope (HST)} when possible to best constrain the predicted gamma-ray flux.  Including this and previous works, we now find the IC/CMB model to be ruled out in a total of 24/45 MSC X-ray jets due to its over-prediction for the observed MeV-to-GeV gamma-ray flux.  We present additional evidence against the IC/CMB model, including the relative X-ray-to-radio relativistic beaming in these sources, and the general mismatch between radio and X-ray spectral indexes. Finally, we present upper limits on the large-scale bulk-flow Lorentz factors for all jets based on the Fermi upper limits, which suggest that these jets are at most mildly relativistic.

\end{abstract}




\section{Introduction}
\label{intro}

The launch of the \textit{Chandra} X-ray observatory in 1999 ushered in a new era of high-resolution, sensitive X-ray imaging of extragalactic jets in active galactic nuclei (AGN). Prior to its launch, it was not clear that high levels of X-ray emission would be a common occurrence in the large-scale (i.e., kpc-scale) AGN jets. However, there are now over $\sim$150 known, resolved X-ray jets, most of which were discovered by \emph{Chandra}\footnote{see e.g. the XJET database http://hea-www.harvard.edu/XJET/ and the ATLASX catalog}.  The X-ray emission in many of the lower-power jets is usually ascribed to synchrotron radiation from the high-energy extrapolation of the electron population producing the radio-to-optical synchrotron spectrum \citep[e.g.,][although see \citealt{meyer+18} for a counter-example]{wilson2002}.  Conversely, a single synchrotron spectrum extending from radio to X-ray wavelengths is precluded in many of the more powerful quasar-hosted jets \citep[e.g.,][]{sambruna2004,jorstad2006,tavecchio2007,kharb+12,stanley15}, where the observed X-ray spectrum is either too high in flux or hard to be consistent with the high-energy extrapolation of the electron energy distribution (EED) producing the radio-to-optical synchrotron spectrum.  In some cases there appear to be at least \emph{three} distinct emission components \citep[e.g., the jet in M84;][]{meyer+18}, and a second component can be seen emerging in the optical/UV in some cases \citep{cara2013,breiding+17}. In \cite{breiding+17} we began referring to jets with one or more additional high-energy emission components beyond the radio synchrotron emission as  ``Multiple Spectral Component'' (MSC) jets, and we will continue using this naming convention for the present study.

The first kpc-scale MSC jet to be detected in X-rays by \emph{Chandra} was hosted by the quasar PKS~0637-752 \citep{chartas2000,schwartz2000}. Soon after, \cite{tavecchio2000} and \cite{celotti2001} independently proposed that the high level of X-ray emission could be explained as inverse-Compton (IC) scattering of the cosmic microwave background (CMB) by the same electron population producing the radio-to-optical synchrotron spectrum (i.e., IC/CMB).  Subsequently, IC/CMB has become the dominant explanation for the anomalously bright and/or hard X-rays seen in MSC X-ray jets \citep[e.g.,][]{sambruna2004,sambruna06,perlman+11,marshall+18,schwartz+20}.  However, due to the very high level of X-ray emission, the IC/CMB model requires high Doppler factors, which in turn requires jets that remain highly relativistic on kpc scales (with bulk-flow Lorentz factors of $\Gamma\sim10$) and closely aligned to the line of sight (leading to Mpc-scale deprojected jet lengths in several cases, \citealt{dermer2004}). The IC/CMB model also requires an extension of the (assumed) power-law electron distribution to very low energies, resulting in high total power, exceeding the Eddington limit in many cases \citep{dermer2004}. 

One strong line of evidence against the IC/CMB model arises from the morphological comparison of the associated X-ray, radio, and optical jet emission components.  In particular, the IC/CMB model requires cospatial emission in these different bands, since the radio-to-optical synchrotron spectrum is produced by the same EED upscattering the CMB to X-ray energies.  However, many MSC X-ray jets contain anomalous X-ray-jet emission components spatially offset from the associated synchrotron radio knots by $\sim$1~kpc in projection; where in most cases the X-ray centroids are located upstream (i.e., closer to the black hole) from the radio centroids \citep[e.g.,][]{karthik+21,karthik+22}.  This is in clear conflict with the expectations from the IC/CMB model, where we would expect if any offset to occur, for the X-ray emission to persist downstream (i.e., further from the black hole) of the peak radio emission due to the longer cooling times of the low-energy portion of the EED responsible for the IC/CMB X-rays.

These long cooling times would also seem to predict X-ray emission which is continuous along the length of the jet, but in most cases the observed X-ray emission is confined to compact knots of order $\sim$1~kpc in size (see \citealt{harris2006} for a thorough review of these issues).  This problem can be avoided by requiring the plasma to be contained within discrete blobs \citep[though see][where this scenario has been ruled out in 3C~273 by the lack of expected proper motions]{meyer+16}.  The radio synchrotron emission from the low-energy electrons (i.e., $\gamma~\sim~100$) have been searched for with the high angular-resolution Low Frequency Array (LOFAR) observations at 150~MHz presented in \cite{harris+19} for the MSC X-ray jet hosted by the quasar 4C~+19.44. However, it was found that the number density of these low-energy electrons was much less than the implied value from the low-energy extrapolation of the GHz-emitting electrons due to the very faint detections of the jet features at 150~MHz (suggesting significant spectral curvature in this jet).  This result is even more problematic for applying the IC/CMB model in this source, since it necessarily implies a unrealistically high level of relativistic beaming in order to account for the anomalous X-rays. 

Other observations which are challenging for the IC/CMB model include the high degree of polarization measured in the rising UV component (clearly part of the second spectral component) in the jet hosted by quasar PKS~1136-135 \citep[][where the IC/CMB radiation is expected to be unpolarized]{cara2013}, a detected X-ray counter-jet in Pictor~A \citep[][where the resulting jet-to-counter jet flux ratio implies a non-relativistic jet oriented at large angles to the line of sight]{hardcastle+16}, flux variability on the order of a few years in the X-ray knots of Pictor~A \citep{hardcastle+16}, rising UV spectral components that have an implied spectral index inconsistent with the IC/CMB model in PKS~2209+080 \citep{breiding+17}, and a lack of correlation between the ratio of X-ray-to-radio jet luminosity with redshift in large samples of X-ray jets \citep[e.g.,][]{marshall+18}.  This latter result is especially challenging, since the CMB energy density is directly proportional to $\mathrm{(1+z)^{4}}$, so the naive expectation is that we would expect more luminous X-ray jets and a higher proportion of X-ray jets with increasing redshift \citep{schwartz02}.  

\begin{table*}
	\centering
	\caption{Sample Properties} 
	\label{tab:example_table}
	\begin{threeparttable}
		\begin{tabular}{llccccccccr} 
			\hline
			Source & Common & & &  &  &   &  &&  &  $\mathrm{IC/CMB}$  \\
			Name & Name & z & kpc/$\arcsec$ & $\mathrm{log\frac{M_{BH}}{M_{\odot}}}$ & log~$\mathrm{L_{kin}}$ & log~$\mathrm{R_{CE}}$ & log~$\mathrm{L_{ext}}$&$\mathrm{B_{eq}}$ & log~$\mathrm{R_{x}}$ & Ruled\\
			(J2000) &  &  &  &  & ($\mathrm{erg\ s^{-1}}$) &&($\mathrm{erg\ s^{-1}}$) & ($\mathrm{10^{-4}~G}$) &  & Out?\\
			\hline
			J0038$-$0207\tnote{$\dagger$} & 3C~17 & 0.220 & 3.7 & ... & 45.4 & -1.55 &42.9& 0.716 & -0.337 & Y\\
			J0108+0135\tnote{$\dagger$} & PKS~B0106+013 & 2.11 & 8.5 & ... & 46.4 & -0.42 &44.4&  2.16& 0.522 & N\\
			J0209+3547 & 4C~35.03 & 0.037 & 0.76 & ... & 43.9 & -1.55 &40.6& 0.782 & 0.846 & Y\\
			J0210$-$5101\tnote{$\dagger$} & PKS~0208$-$512 & 1.0 & 8.2 & ... & 46.0 & -0.52  &43.8& 0.422 & 0.826 & Y\\
			J0237+2848\tnote{$\dagger$} & 4C~28.07 & 1.21 & 8.5 & ... & 45.7 & 0.42  &43.4& 1.12 & 0.649 & Y\\
			J0416$-$2056 & PKS~0413$-$210 & 0.808 & 7.7 & 8.2$\mathrm{^{c}}$ & 45.9 & -0.29  &43.7& 1.51 & -0.40 & N\\
			J0418+3801\tnote{$\dagger$} & 3C~111 & 0.05 & 1.0 & ... & 44.7 & -0.52  &41.8& 0.423 & 1.77 & Y\\
			J0433+0521\tnote{$\dagger$} & 3C~120 & 0.034 & 0.70 & 7.8$\mathrm{^{e}}$ & 43.9 & 0.43 &40.6& 0.322 & 1.64 & Y\\
			J0519$-$4546\tnote{$\dagger$} & Pictor~A & 0.034 & 0.70 & 7.6$\mathrm{^{b}}$ & 44.9 & -1.98 &42.1& 0.239 & 1.28 & Y\\
			J0607$-$0834\tnote{$\dagger$} & PKS~0605$-$085 & 0.870 & 7.9 & ... & 45.2 & 1.04 &42.6&0.680 & 0.907 & Y\\
			J0728+6748 & 3C~179 & 0.844 & 7.9 & 8.6$\mathrm{^{c}}$ & 45.9 & -1.15&43.7& 0.740 & 0.543 & Y\\
			J0741+3112 & B2~0738+31 & 0.631 & 7.0 & 9.3$\mathrm{^{c}}$ & 45.0 & 0.56 &42.4& 0.138 & 1.64 & N\\
			J0830+2410\tnote{$\dagger$} & B2~0827+24 & 0.941 & 8.1 & 8.4$\mathrm{^{h}}$ & 45.3 & 0.42 &42.8& 0.309 & 1.37 & Y\\
			J0840+1312\tnote{$\dagger$} & 4C~13.38 & 0.680 & 7.3 & 8.5$\mathrm{^{e}}$ & 45.9 & -1.06 &43.6& 1.67 & 0.531 & Y\\
			J0922$-$3959\tnote{$\dagger$} & PKS~0920$-$397 & 0.590 & 6.8 & ... & 45.6 & -0.26 &43.4& 0.813 & 0.199 & Y\\
			J0947+0725 & 3C~227 & 0.086 & 1.7 & ... & 44.9 & -2.75  &42.1& 0.185 & 0.052 & N\\
			J1001+5553 & QSO~0957+561 & 1.41 & 8.7 & 9.4$\mathrm{^{a}}$ & 45.9 & -1.84 &43.7& 1.66 & -0.244 & N\\
			J1007+1248 & 4C~13.41 & 0.241 & 3.9 & 9.3$\mathrm{^{e}}$ & 45.0 & -2.21 &42.3& 0.103 & -0.315 & N\\
			J1033$-$3601 & PKS~1030$-$357 & 1.46 & 8.7 & ... & 46.1 & -0.41  &44.0& 1.02 & 0.628 & N \\
			J1048$-$1909 & PKS~1045$-$188 & 0.590 & 6.8 & 6.8$\mathrm{^{e}}$ & 45.5 & -0.52 &43.0& 0.90 & 0.138 & N\\
			J1048$-$4114 & PKS~1046$-$409 & 0.620 & 7.0 & ... & 45.7 & -1.52 &43.4& 0.834 & 0.180 & Y\\
			J1058+1951 & 4C~20.24 & 1.11 & 8.4 & ... & 46.1 & -0.39  & 44.0&0.469 & 0.302 & Y\\
			J1130$-$1449\tnote{$\dagger$} & PKS~1127$-$145 & 1.19 & 8.5 & ... & 45.6 & 1.28 &43.3&0.150 & 0.794 & N\\
			J1153+4931\tnote{$\dagger$} & PKS~1150+497 & 0.334 & 4.9 & 8.6$\mathrm{^{c}}$ & 45.1 & 0.19 &42.5& 0.742 & 1.27 & Y\\
			J1205$-$2634\tnote{$\dagger$} & PKS~1202$-$262 & 0.786 & 7.7 & 8.8$\mathrm{^{c}}$ & 45.8 & -0.61 &43.5& 1.13 & 1.03 & Y\\
			J1224+2122\tnote{$\dagger$} & 4C~21.35 & 0.434 & 5.8 & 8.5$\mathrm{^{d}}$ & 45.4 & -0.79 &42.9& 0.974 & 0.802 & N\\
			J1319+5148 & 4C~52.27 & 1.06 & 8.3 & 9.5$\mathrm{^{d}}$ & 45.9 & -0.54  &43.6& 1.03 & 0.662 & N\\
			J1325+6515 & 4C~65.15 & 1.63 & 8.7 & 9.6$\mathrm{^{a}}$ & 46.1 & -1.30  &44.0& 4.0 & -0.265 & N\\
			J1421$-$0643 & PKS~1418$-$064 & 3.69 & 7.4 & ... & 46.0 & -0.63  &43.8& 1.26 & 1.92 & N\\
			J1510+5702\tnote{$\dagger$} & 
			QSO~B1508+572  & 4.31 & 6.9 & ... & 45.8 & -0.15  & 43.6&0.526 & 1.38 & Y\\
			J1632+8232\tnote{$\dagger$} & NGC~6251 & 0.020 & 0.42 & 8.8$\mathrm{^{g}}$ & 43.9 & -0.83  &40.5& 0.148 & -0.575 & N\\
			J1642+3948\tnote{$\dagger$} & 3C~345 & 0.595 & 6.9 & 9.1$\mathrm{^{h}}$ & 45.8 & 0.33  &43.5& 1.50 & 0.075 & N\\
			J1642+6856 & 4C~69.21 & 0.750 & 7.6 & 7.3$\mathrm{^{c}}$ & 45.5 & 0.03  &43.1& 0.709 & 0.051 & Y\\
			J1720$-$0058 & 3C~353 & 0.030 & 0.63 & ... & 44.8 & -2.59  &42.0& 0.348 & 0.365 & Y\\
			J1746+6226 & 4C~62.29 & 3.89 & 7.2 & ... & 46.5 & -0.51 &44.7 & 3.19 & 1.33 & N\\
			J1829+4844\tnote{$\dagger$} & 3C~380 & 0.692 & 7.3 & 9.8 & 46.4 & -0.64 &44.4& 2.44  & 0.089 & Y\\
			J1849+6705\tnote{$\dagger$} & 8C~1849+670 & 0.660 & 7.2 & 9.1$\mathrm{^{e}}$ & 45.0 & 0.51  &42.3& 0.226 & 1.70 & N\\
			J1927+7358 & 4C~73.18  & 0.30 & 4.6 & 8.7$\mathrm{^{e}}$ & 45.0 & 0.50 &42.2& 0.520 & 2.20 & N\\
			J2005+7752\tnote{$\dagger$} & S5~2007+777 & 0.342 & 5.0 & 7.4$\mathrm{^{f}}$ & 44.4 & 1.18  &41.4& 0.337 & 1.03 & N\\
			J2105$-$4848 & PKS~2101$-$490 & 1.04 & 8.3 & ... & 45.8 & -0.06  &43.5& 0.589 & 0.105 & N\\
			J2158$-$1501\tnote{$\dagger$} & PKS~2155$-$152 & 0.670 & 7.2 & 7.4$\mathrm{^{c}}$ & 45.4 & 0.30  &43.0& 0.975 & 0.066 & N\\
			J2203+3145 & 4C~31.63 & 0.295 & 4.5 & 8.9$\mathrm{^{e}}$ & 44.9 & 0.03 &42.2& 0.199 & 0.919 & N\\
			J2218$-$0335 & PKS~2216$-$038 & 0.901 & 8.0 & 9.2$\mathrm{^{c}}$ & 45.6 & -0.47  & 43.3&0.572 & 0.028 & N\\
			J2253+1608\tnote{$\dagger$} & 3C~454.3 & 0.859 & 7.9 & 8.9$\mathrm{^{c}}$ & 45.8 & 0.79  &43.6& 1.38 & 0.877 & N\\
			J2338+2701 & 3C~465 & 0.030 & 0.62 & ... & 44.3 & -1.63  &41.1&  0.864 & 0.620 & Y\\
			\hline
		\end{tabular}
		\begin{tablenotes}
			\item Black hole mass measurements are obtained from the following literature sources (corresponding to the superscripts in the $\mathrm{M_{BH}}$ column): (a) \cite{kozlowski17}, (b) \cite{lewisanderacleous06}, (c) \cite{liu2006}, (d) \cite{wang+04}, (e) \cite{wooandurry02}, (f) \cite{wu+02}, (g) \cite{vandenbosch16}, (h) \cite{xie+05}.
			\item[$\dagger$]These sources are members of the 4FGL catalog; for these sources we used the re-combined light curve analysis described in section~\ref{fermi_methods} to determine the upper limits/minimum flux levels appropriate for the steady-state large-scale-jet gamma-ray flux.
		\end{tablenotes}
	\end{threeparttable}
\end{table*}

\begin{table*}
	\begin{threeparttable}
	\caption{Radio Observations} 
	\label{radio-table}
	\begin{tabular}{l  c c c c c c c r}
		\hline
		\centering
		Source&Telescope&Observing&Observing&&&&\\
		Name &Array&Band & Frequency & Project Code & Observation Date & RMS & Beam Size & P.A\\
		(J2000)&&& (GHz) & & YYYY-MM-DD & (mJy beam$^{-1}$) & ($^{\prime\prime}~\times~^{\prime\prime}$) & ($^{\circ}$)
		\\ 
		\hline
		J0038$-$0207&VLA&C&4.86&AS0179&1985$-$03$-$08\tnote{$\dagger$}&0.870&0.49$\ \times\ $0.39&23.4\\
		J0108+0135&VLA&C&4.86&AR197&1989$-$01$-$07\tnote{$\dagger\dagger\dagger$}&0.234&0.47~$\times$~0.40&19.7\\
		J0108+0135&VLA&X&8.44&AR197&1989$-$01$-$08\tnote{$\dagger\dagger$}&0.123&0.27~$\times$~0.23&10.5\\
		J0108+0135&VLA&Ku&15&AR223&1990$-$04$-$29\tnote{$\dagger\dagger$}&0.520&0.20~$\times$~0.15&57.9\\
		\hline \\
	\end{tabular}
\begin{tablenotes}
	\item This table is published in its entirety online in machine-readable format, where we show a sample portion here for guidance regarding its form and content.
	\item[$\dagger$]These observations were used for making the images displayed in Figures~\ref{fig:img1}-\ref{fig:img3}.
	\item[$\dagger\dagger$]These observations were used for flux density measurements.
	\item[$\dagger\dagger\dagger$]These observations were used for both the images displayed in Figures~\ref{fig:img1}-\ref{fig:img3} and flux density measurements.
\end{tablenotes}
\end{threeparttable}
\end{table*}

Despite all of the weight of evidence above, it is still possible that the IC/CMB mechanism is the dominant emission process at work in some resolved, kpc-scale X-ray jets. This may in particular be the case for jets at very high redshift due to the strong CMB enhancement \citep[e.g.][]{worrall2020,schwartz2020,ighina2022} or at low redshift, in rare cases of fast kpc-scale jets with favorable alignment \citep{meyer+19}.  In order to evaluate the viability of the IC/CMB model as the dominant X-ray emission mechanism in any individual jet we must test the various predictions it makes for the physical properties of these jets and the resulting multi-wavelength radiation they produce.

One such prediction of an IC/CMB origin for the X-ray emission in resolved X-ray jets is the requirement of a high level of MeV-to-GeV gamma-ray flux \citep[][]{georganopoulos2006}.  This requirement results from the fact that the IC/CMB spectrum is essentially an exact copy of the radio-to-optical synchrotron spectrum, shifted in frequency and luminosity by an amount necessary to reproduce the observed X-ray flux.  This shift is parametrized by B/$\delta$, where B is the magnetic field strength of the emission region and $\delta$ is the Doppler factor\footnote{The Doppler factor,$\delta$, is given by $\mathrm{\delta=\frac{1}{\Gamma\left(1-\beta cos\theta\right)}}$.  Here, $\Gamma$ is the bulk-flow Lorentz factor, $\beta$ is
the bulk flow speed scaled by c, and $\theta$ is the jet angle to the line of sight.}.  There are no other free parameters in the model, although we must extrapolate from the observed $\sim$~GHz radio frequencies to the much lower unobserved tens-to-hundreds of MHz frequencies.  

In several previous studies we have tested the prediction of a high GeV flux using observations from the \textit{Fermi}/LAT gamma-ray space telescope.  
Flux upper limits resuling from the analyis of this \textit{Fermi}/LAT data have ruled out the IC/CMB model in the quasars 3C~273 \citep{meyer+14}, PKS~0637-752 \citep{meyer+15,meyer+17}, and four MSC X-ray jets hosted by quasars not detected by \textit{Fermi} \citep{breiding+17}: PKS~1136-135, PKS~1129-021, PKS~1354+195, and PKS~2209+080.  
In this paper,  we present results from the analysis of over 10 years of \emph{Fermi}/LAT observational data for an additional 45 extragalactic X-ray jets. Of these, 36 are clearly MSC jets based on the X-ray flux and/or spectral index. The remaining nine jets do not have enough constraints to rule out a single-synchrotron spectrum from radio to X-rays but have been modeled as IC/CMB in the literature previously.  
Importantly, in this work we utilize multi-wavelength data obtained from a number of archival and proprietary observations in order to better constrain the radio-to-optical synchrotron spectrum for our sample $-$ leading to a better characterization of the predicted IC/CMB spectrum in the \emph{Fermi}/LAT band.

In section~\ref{methods} we further describe the multi-wavelength data used along with the analysis methodologies utilized.  In section~\ref{results}, we present the results of these analyses, with their interpretations and implications discussed in section~\ref{discussion}.  Finally, in section~\ref{conclusion} we summarize the main findings of our work.  Throughout this paper we adopt a $\Lambda$CDM cosmology, with H$_{0}=67.74$ km s$^{-1}$ Mpc $^{-1}$, $\Omega_{\lambda}=0.69$, and $\Omega_{m}=0.31$ \citep{planck+16}.  Additionally, we use the following spectral index convention: $\mathrm{F_{\nu}\propto\nu^{-\alpha}}$, where $\mathrm{F_{\nu}}$ is the flux density and $\nu$ is the observer-frame frequency of the radiation.  

\pagebreak
\section{Data Analysis}
\label{methods}
\noindent \textbf{Radio Interferometric Data}\\ 

New and archival data from the Very Large Array (VLA),  Australia Telescope Compact Array (ATCA), and the Atacama Large Millimeter/submillimeter Array (ALMA) were reduced for this study.  All radio interferometric data were reduced using the Common Astronomy Software Applications (\textit{CASA}) package \citep{mcmullin+07}.  Below we describe the observations used from each telescope array, and the procedures utilized to calibrate and image the visibility data.  We describe all of the radio interferometric data used in this paper in Table~\ref{radio-table}, where we list the target source name, central frequency and band of the observation, project code, observation date, telescope array, root mean square (RMS) intensity noise level of the final image, and the beam width/position angle (P.A) of the synthesized beams.  
For any flux densities measured for jet components, we assume very conservative 35\% errors on these measured flux densities.  These errors assume a level of $\sim$20\% error on absolute flux calibration and $\sim$15\% error associated with comparing literature flux densities with our own measured flux densities at different frequencies in the same SED.  These conservative error estimates also allow us to account for potential minor differences in flux extraction regions at different bands and for loss of flux associated with spatial frequency filtering inherent in radio interferometric data with different array configurations and observing frequencies.

\subsection{Very Large Array}

We reduced 77 archival observations taken with the VLA for the present study. We also analyzed six new VLA observations taken for this project under the program 15A-357.  
For historical VLA observations before the upgrade to the Karl G. Jansky Very Large Array (JVLA) in 2011, 
we followed a standard data reduction procedure for continuum data. For most observations, 3C~286, 3C~48, or 3C~147 served as the flux calibrator and in most cases the source itself served as the phase calibrator. For more recent JVLA observations, we used the \textit{CASA} pipeline for initial calibration.  Post calibration, any excess radio frequency interference (RFI) was removed and visibility data was imaged using the CLEAN deconvolution algorithm, via the \textit{CASA} task \texttt{tclean}.  Typically, the images were made with Briggs weighting and robust parameter of 0.5 (although sometimes a natural weighting was utilized).  When appropriate or essential to our science goals for the wide bandwidth JVLA data, we used a multi-term, multi-frequency synthesis (mtmfs) deconvolution and imaging procedure (with \texttt{nterms=2}) defined in the \texttt{tclean} task in \textit{CASA}. This type of analysis relies on modeling the sky intensity distribution as a Taylor polynomial, expanded about some reference frequency for the subsequent deconvolution and image reconstruction \citep{rau_cornwell_2011}. In all cases, we applied several rounds of (non-cumulative) phase-only self calibration, followed by a single round of amplitude and phase self-calibration to the data to improve the final image.

\subsection{Atacama Large Millimeter/submillimeter Array (ALMA)}
We analyzed 11 archival observations from ALMA for this project, and 17 new observations from our projects 2015.1.00932.S, 2016.1.01481.S, and 2017.1.01572.S. The details of the observations are given in Table \ref{radio-table}.  The continuum visibility data were calibrated with the \textit{CASA} pipeline via the NRAO-provided python script \texttt{scriptForPI.py}, which gives a science-ready measurement set. Similar to the VLA data, we then imaged the visibility data via the \textit{CASA} task \texttt{tclean} (with Briggs weighting and robust parameter of 0.5).  MTMFS imaging was employed with \texttt{nterms=2}, and typically several rounds of phase-only self calibration and a single round of amplitude and phase self-calibration were applied to further reduce the RMS of the final image.

\subsection{Australia Telescope Compact Array (ATCA)}
For this project, we analyzed six observations from ATCA. 
\textit{MIRIAD} was used to initially load the visibility data and then export into a form to be read by \textit{CASA} for further calibration and imaging.
Standard calibration suitable for continuum observations was applied to the data after flagging any potential RFI in \textit{CASA}. For all ATCA observations used in this project, 1934-638 served as the flux calibrator, and bright sources served as their own phase calibrator. Imaging deconvolution was accomplished using the \textit{CASA} task \texttt{clean} with Briggs weighting and a robust parameter of 0.5. In all cases at least one round of phase-only and a final round of amplitude and phase self-calibration was used to improve the images.

\subsection{\textit{Hubble Space Telescope (HST)}  }

We analyzed \textit{Hubble Space Telescope} (HST) archival observations for the quasars 4C 31.63, QSO 0234+285, PKS~0920-397, PKS~1045-188, and PKS~2155-152.  For all of these galaxies, the flat-field-corrected images were first downloaded from the Mikulski Archive for Space Telescopes (MAST) after having first been processed through the \textit{calwf3} data reduction pipeline.  For 4C 31.63, we obtained a 1.1 ks WFPC2/PC1 (F555W: 544~nm) observation from project U2CY0 which contained only a single frame.  For PKS~1045-188, we analyzed two 2.7~ks dithered observations taken with the WFC3 detector in the IR and UV channels (filters F160W/F475W and poject code IC6701).  For the remaining quasars, we analyzed dithered observations taken with the ACS/WFC1 detector and F814W filter (805~nm) and ranging from 2 to 2.3~ks in length (project codes: JC0R02, JC0R02, and J9740).  As all observations show a bright point source at the jet `core', we performed a point spread function (PSF) subtraction to obtain a measurement of the extended jet.  For this purpose, we used the publicly available \textit{GALFIT} \citep[GALaxy-FITting;][]{peng+10} software package.  All model-fitting was done in the non-distortion-corrected image frames, where we first masked pixels corresponding to other sources, cosmic ray pollution, and other image artifacts prior to the fitting. The host galaxies were each fit with single S\'ersic components and we used nearby stars, \textit{TinyTim}, or a dedicated PSF star observation for the PSF models depending on which model best-fit the observed quasar point source response.  Finally, we subtracted the best-fit PSF and S\'ersic models from the images and combined the dithered PSF-subtracted image files together with  \textit{Astrodrizzle} \citep{gonzaga+12}, which also subtracts the sky background and makes corrections for cosmic ray artifacts and geometric distortion.  All of the jet components for these quasars remained undetected in the \textit{HST} imaging except for PKS~0920-397 and PKS~1045-188.  We determined jet component flux density upper limits for the non-detections based upon average count rate RMS deviations in regions the same distance from the core as the jet knots (identified from radio imaging).  For the detected jet components, we also used this method to determine the background RMS deviations to be combined with the Poisson count noise when determing our photometric errors.  We used the inverse flux sensitivities and pivot wavelengths provided in the FITS file headers in order to determine the conversion from count rate to flux densities.  Appropriate extinction corrections were applied \citep{schlafly+11}, along with the relevant aperture corrections \citep{bohlin+16}.

\subsection{Chandra}
We re-analyzed a number of archival \emph{Chandra} imaging observations in order to compare radio and X-ray morphologies for our sample, and to obtain X-ray spectral index measurements for the relevant emission regions.  We reduced the X-ray data using the standard methods described in the Chandra Interactive Analysis of Observations \citep[CIAO ][]{2006SPIE.6270E..1VF} threads. Briefly,  we performed the initial re-processing with the \texttt{chandra\_repro} task in CIAO using calibration data from CALDB v4.9.4 and discarded events outside the range of 0.4-8 keV. We examined each light curve and excluded events observed during periods with count rates above 2$\sigma$ above the mean level. For sources with multiple observations, each remaining observation was aligned to the one with maximum exposure by matching the centroids of their cores and were merged using the \texttt{merge\_obs} task. Finally, we binned the final events file by a factor of 0.2 (1/5th native ACIS-S pixel) to produce a sub-pixel image for each source.

We extracted the spectra for each source using the \texttt{spectract} task in CIAO and fit an absorbed power-law model in SHERPA \citep{2001SPIE.4477...76F} for events falling in the 0.4-8 keV range; the hydrogen absorption column density was fixed to the reference value based on sky location\footnote{Retrieved using WebPIMMS service (\url{https://heasarc.gsfc.nasa.gov/cgi-bin/Tools/w3nh/w3nh.pl})}. We binned the spectra to 1 ct/bin and used Nelder-Mead optimization method with WStat statistic to perform the fitting.  Further details concerning the X-ray data analysis and measurements for each jet will be given in an upcoming publication \citep{karthik+22}.

\subsection{Fermi}
\label{fermi_methods}

\begin{figure}
	\includegraphics[scale=0.45]{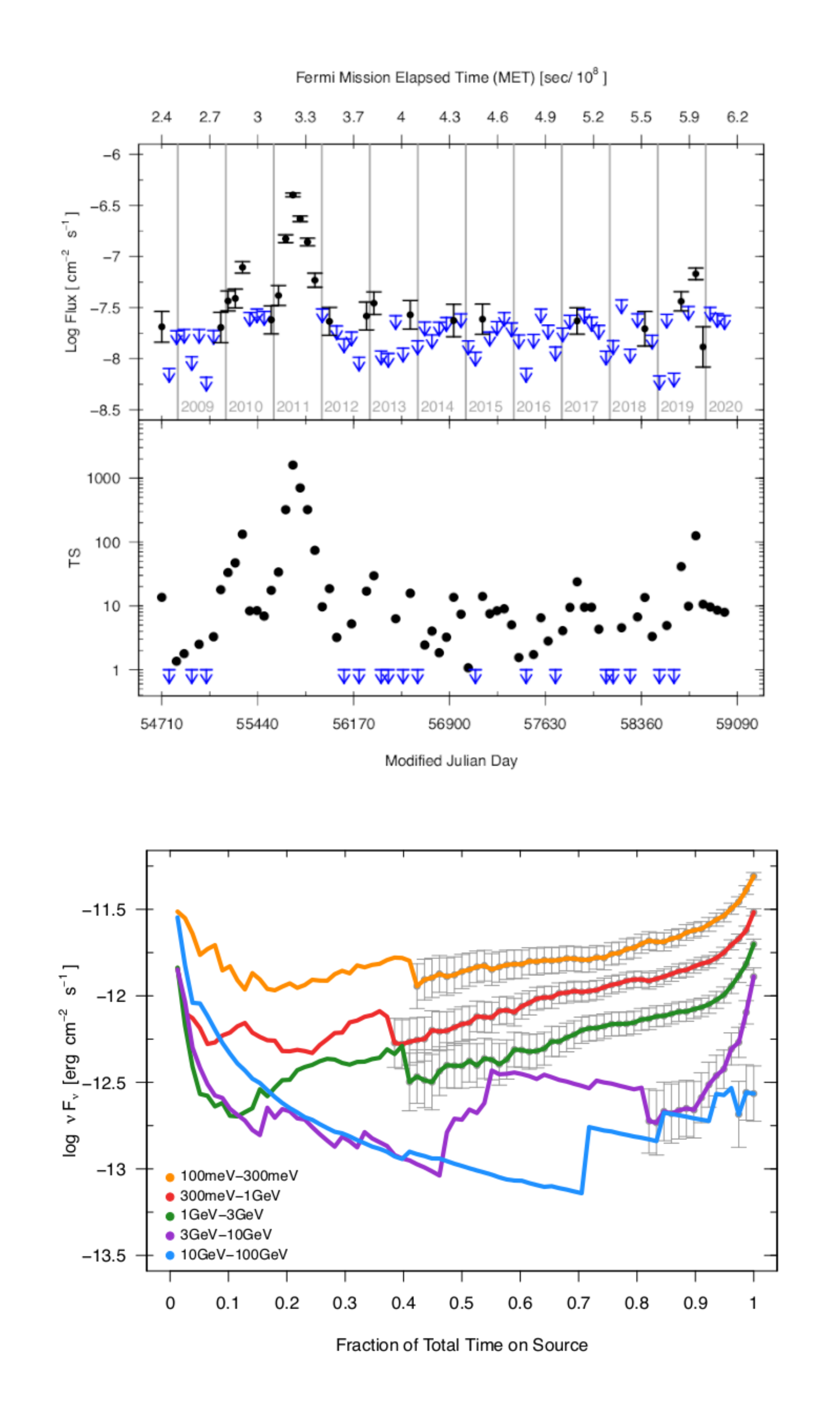}
	\caption{Top: \textit{Fermi}/LAT light curve (upper panel) for PKS~1150+497 with corresponding TS values for each bin shown in the lower panel. The time bins correspond to three weeks in good time interval, and the flux from the source is the integrated photon flux from 100~MeV$-$100~GeV. The blue arrows in the light curve represent upper limits for bins in which the source TS was less than ten. Black data points represent fluxes measured when the TS was greater than ten, with 1$\sigma$ error bars. In the TS plot, blue arrows represent TV values less than zero, while the black points represent positive TS values. Bottom: Combined-bin flux density upper limits for PKS~1150+497 are shown as lines without error bars, where upper limits are measured when the source TS was less than 10. Flux densities are shown with gray 1$\sigma$ error bars on top of the lines when the source TS was greater than ten. The flux densities/upper limits are given in the five Fermi energy bands described in section~\ref{fermi_methods}.}
	\label{fig:lc}
\end{figure}

\begin{figure*}
	\includegraphics[scale=0.51]{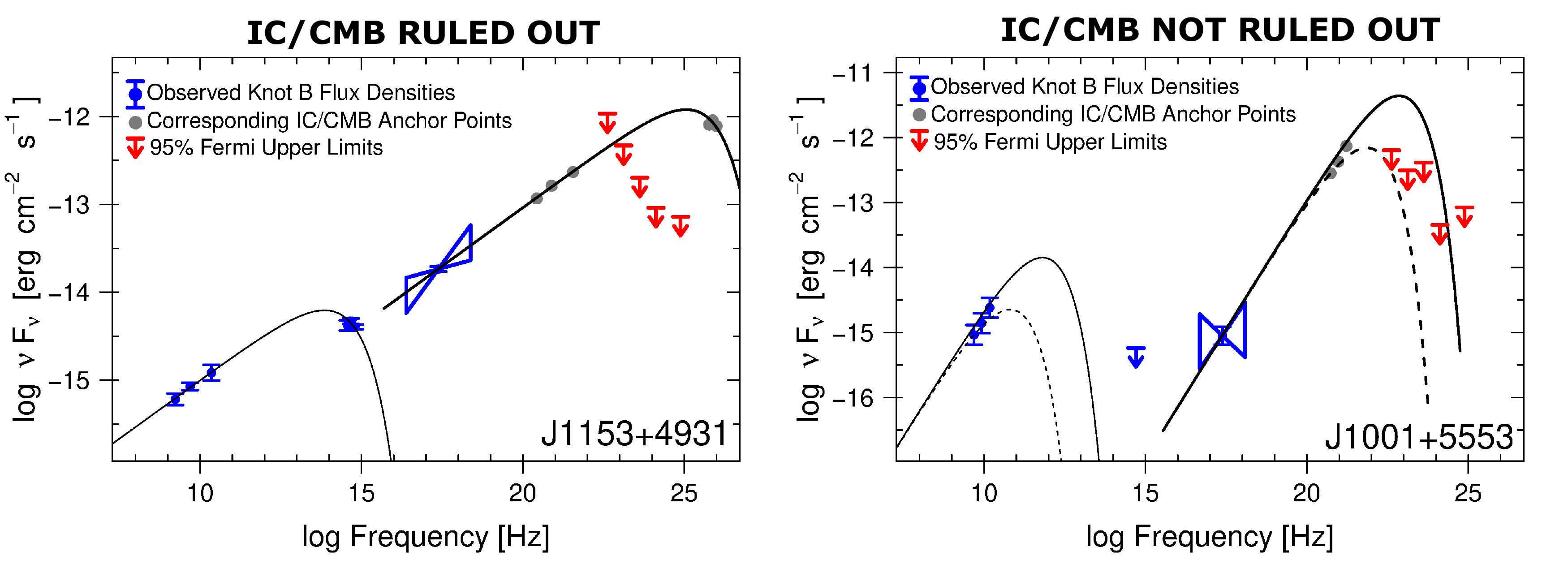}
	\caption{ Here we show two example SEDs for the X-ray-jet features analyzed in this study, hosted by the quasars PKS~1150+497 (left) and QSO~0957+561 (right).  These SEDs are illustrative of the two possible outcomes of applying our test of the IC/CMB model using \textit{Fermi}/LAT observations: namely whether it is ruled out or not based upon the predicted level of MeV-to-GeV gama-ray flux.  Observed flux densities are shown with blue circles (plus 1$\sigma$ error bars), and bowties represent uncertainties in the measured X-ray spectral indexes.  Downward-facing arrows represent upper limits based upon non-detections, where we show the gamma-ray upper limits determined by \textit{Fermi}/LAT observations in red.  We show the phenomenological synchrotron fits as thin black lines and the IC/CMB model fits as thick black lines.  Gray points mark where the measured radio-to-optical synchrotron flux densities would lie on the shifted IC/CMB spectra, and thus represent anchor points for the IC/CMB spectra where there is no uncertainty in the predicted flux densities.  
	For QSO~0957+561 we show two possible synchrotron models which are consistent with the observed radio-to-optical observations as a thin black line and dotted line with corresponding IC/CMB model curves as a thick black line and dashed line, respectively -- see text. 
	The SEDs for the remaining sample can be found in Figures~\ref{fig:sed1} \& \ref{fig:sed2}. }
	\label{fig:sed_showcase}
\end{figure*}
For all of our sources, we used all of the available \textit{Fermi}/LAT data at the time of analysis, ranging from $\sim$~11.7 to 12.1 years (mission elapse end times for the extracted data can be found in Table~\ref{fermi-table}).  \textit{Fermi}/LAT event and spacecraft data were extracted using a 10$^{\circ}$ region of interest (ROI), an energy cut of 100~MeV$-$100~GeV, a zenith angle cut of $90^{\circ}$, and the recommended event class and type for point source analysis (128 and 3, respectively).  Following the standard methodology for \textit{Fermi}/LAT binned likelihood analysis, a binned counts map was made with 30 logarithmically-spaced energy bins and 0.2 degree spatial bins.  An initial spatial and spectral model file was constructed with sources up to $10^{\circ}$ outside the ROI using the publicly available \texttt{make4FGLxml.py} script, which populates the model file with point and extended sources from the \textit{Fermi}/LAT 4FGL catalog and an extended source catalog, respectively.  Additionally, the galactic diffuse emission model, \texttt{gll\_iem\_v07.fits}, and recommended isotropic diffuse emission model for point sources, \texttt{iso\_P8R3\_SOURCE\_V3\_v1.txt}, were used for the analyses.  The livetime cubes were computed using 0.025 steps in cos$(\theta)$ (where $\theta$ is the inclination with respect to the LAT's z-axis) and $1^{\circ}$ spatial binning.  Then all-sky exposure maps were computed using the same energy binning as the counts map. After obtaining converged fits with the maximum likelihood optimizations, we determined upper limits (or lowest flux states) for the large-scale-jet gamma-ray fluxes from our sources in the five energy bands: 100~MeV$-$300~MeV, 300~MeV$-$1~GeV, 1~GeV$-$3~GeV, 3~GeV$-$10~GeV, and 10~GeV$-$100~GeV.  This procedure is straightforward for sources which are not detected by \textit{Fermi}, and are not members of the 4FGL point source catalog.  In this case, we forced all of the model parameters fixed and added a point source with power-law spectral model at the sky position of our source of interest with normalization of the power law set free and a fixed photon index of 2 (corresponding to a flat spectrum in $\mathrm{\nu F_{\nu}}$).  The 95\% flux density upper limits calculated with the \textit{Fermi} tools then apply to the entire source and not just the large-scale jet.  For sources detected by \textit{Fermi} which are members of the 4FGL catalog, we use the procedure outlined below to determine constraints on the large-scale-jet gamma-ray flux.  
\\

\noindent \textbf{Recombined Lightcurves with the \textit{Fermi}/LAT}
\\

We now describe a method for detecting the level of a steady flux, or deriving an upper limit on such a steady flux, when the signal coincides (or is contaminated by) an unrelated variable component \citep[as first implemented and described by][]{meyer+14}.  We utilize this method for the particular case of determining (or placing limits on) the level of steady gamma-ray flux coming from a kpc-scale AGN jet, where the base of the jet (or 'core', on sub-pc scales near the black hole) is known to be a highly variable and strong gamma-ray emitter (although it should be more widely applicable in principle). The spatial separation of the two emitting regions is on the order of thousands of light-years and thus they are completely (causally) disconnected. Observationally, the angular separation between the variable core and the steady jet is generally on the order of arcseconds (or arcminutes at most), which means that for a telescope like the \emph{Fermi}/LAT, with at best 0.1 degree resolution, these two gamma-ray sources appear co-spatial.  The general principle behind this method is to utilize data from periods when the bright and variable component is in its most quiescent states, allowing for the longest integration times when the core is not producing a high level of gamma-ray emission and one has a better chance of detecting the steady and low level of IC/CMB gamma-ray flux.  

In the present study, we implemented this procedure by first producing a light curve, with data subdivided into three week time bins\footnote{The time bins were defined in terms of good time interval (GTI) time, corresponding to roughly eight weeks in real time.  The time bins are large enough to avoid being in the photon starved regime, so as to steer clear from any potential biases in the time bin recombination analysis (defined in this case as containing at least two photons in any given time bin of the analysis assuming Poisson statistics).} and executing the pre-likelihood analysis tools on each time bin separately.  Then maximum likelihood optimizations were computed for the data from each time bin to obtain best-fit normalizations for all sources in the model file, with the other spectral parameters fixed.  Subsequently, the Test Statistic\footnote{The Test Statistic is formally defined as $-2~\times$ the natural logarithm of the ratio of the maximum likelihood of a model without an additional source to a model with an additional source at the specified sky position.} (TS, roughly equivalent to $\sigma^{2}$) of our sources were computed for each time bin, and we recombined the data from lowest to highest TS (one time bin at a time).  Next, we re-ran the pre-likelihood analysis tools on the data from each contiguous time bin addition, and obtained converged maximum likelihood fits.  After this, we force all of the parameters in the model file fixed except those of the source of interest before computing flux upper limits (or fluxes if the TS of the source was greater than 10) for each energy band.  The resulting upper limits (or fluxes) apply to the entire source including the IC/CMB gamma-ray flux produced by the large-scale jet.  Thus, the gamma-ray limits (or minimum fluxes) obtained from this analysis apply to the gamma-ray emission from the entire jet, not just any individual knot or jet component.

In Figure~\ref{fig:lc}, we show an example \textit{Fermi}/LAT light curve and corresponding plot of the upper limits/flux densities with each time bin addition for the source PKS~1150+497 (light curves and combined bin plots are available as online-only figures). As is readily apparent, the upper limits decrease with added integration time, until time bins corresponding to increased flux levels are added into the analysis and the upper limits begin to increase, eventually becoming increasingly bright detections.  One key feature we would expect to see for any IC/CMB detection in gamma-rays using this method is a plateau in flux with decreasing error bars, before photons from the bright and variable core dominate the flux and the observed fluxes begin to increase again \citep[see, e.g.,][]{meyer+19}.  For any recombined light curve analysis where we present a flux density instead of an upper limit in any given band for the minimum flux level consistent with the steady-state gamma-ray flux, and the recombined light curve does not have the above-mentioned properties, this flux density likely represents the minimum \textit{core flux} over the length of the \textit{Fermi}/LAT observations.  The IC/CMB gamma-ray flux must be at or below this level, so any use of these flux minima would represent conservative estimates when considering whether or not the IC/CMB model for the large-scale-jet X-ray emission is consistent with the \textit{Fermi}/LAT observations.

This methodology relies on the assumption of a bright and variable core superposed with a weaker, steady state IC/CMB jet component.  We discuss the robustness of the constraints on any potential IC/CMB component derived from this approach in Appendix~\ref{simulation} with supporting simulations using the \textit{Fermi} science tools.

\section{Results}
\label{results}

The primary goal of this study is to assess the viability of the IC/CMB model accounting for the anomalous X-ray emission in resolved (kpc-scale), MSC X-ray jets hosted by AGN.  To this end, we analyzed data from \textit{Fermi}/LAT observations to search for the predicted high levels of IC/CMB gamma-ray flux.  The expected gamma-ray signature from the IC/CMB model is a non-variable and (usually) hard spectral state, which can be isolated from the soft and highly variable bright quasar core during times when the core is in a quiescent state.  Following the methodology presented in section~\ref{fermi_methods}, we searched for the plateau signature in all 23 of the \textit{Fermi}-detected MSC jets which are members of the 4FGL catalog.  We did not find evidence for such a gamma-ray signature in any of the \textit{Fermi}/LAT observations of the AGN presented here.  Thus, we utilize all of the data during times when the variable core is quiescent in order to obtain the deepest constraints on the large-scale-jet gamma-ray flux.  For the  22 sources not in the 4FGL catalog, we utilize all available \textit{Fermi}/LAT data for the entire time \textit{Fermi} is observing our source of interest in order to obtain the deepest flux density upper limits.  The viability of the IC/CMB model is then assessed on the basis of whether or not the \textit{Fermi}/LAT flux density upper limits (or minimum flux densities) are consistent with predictions from the IC/CMB model.  For cases in which the IC/CMB model over-predicts the level of gamma-ray flux consistent with the  \textit{Fermi}/LAT data, we can rule out the IC/CMB model.  Conversely, if the IC/CMB model curve lies below the minimum \textit{Fermi}/LAT flux density upper limits (or flux densities in the case of source detections), the IC/CMB model can not be ruled out; in the absence of a clear plateau signature we remain agnostic about its validity as the correct model.      

\begin{table*}
	\begin{threeparttable}
		\caption{\textit{Fermi} Analysis Results} 
		\label{fermi-table}
		\begin{tabular}{l c c c c c c c c c r}
			\hline
			\centering
			Source&Observing&Observing&&&Log&Predicted&Observed&Observed&\\
			Name&End Time&Band&E1&E2&Freq&$\mathrm{\nu F_{\nu,IC/CMB}}$& $\mathrm{\nu F_{\nu}}$&$\mathrm{\nu F_{\nu}}$ Error&$\delta$\\
			(J2000)&(MET in s)&&(GeV)&(GeV)&(Hz)&$\mathrm{erg\ s^{-1}\ cm^{-2}}$&$\mathrm{erg\ s^{-1}\ cm^{-2}}$&$\mathrm{erg\ s^{-1}\ cm^{-2}}$&limit\\ 
			\hline
			J1153+4931&617349822&1&0.1&0.3&22.6&$4.06\times10^{-13}$&< $1.08\times10^{-12}$&&< 6.3\\
			&&2&0.3&1&23.1&$5.09\times10^{-13}$&< $4.67\times10^{-13}$&&\\
			&&3&1&3&23.6&$6.07\times10^{-13}$&< $2.01\times10^{-13}$&&\\
			&&4&3&10&24.1&$6.49\times10^{-13}$&< $9.16\times10^{-14}$&&\\
			&&5&10&100&24.7&$5.32\times10^{-13}$&< $7.24\times10^{-14}$&&\\
			\hline \\
		\end{tabular}
		\begin{tablenotes}
			\item This table is published in its entirety online in machine-readable format, where we show a sample portion here for guidance regarding its form and content.
		\end{tablenotes}
	\end{threeparttable}
\end{table*} 
In Figure~\ref{fig:sed_showcase} we show two representative example SEDs from our sample to compare the cases in which we can confidently rule out the IC/CMB model to those in which we can not (the remaining jet SEDs can be found in Appendix~\ref{images_SEDs}).  In these SEDs we show observed flux densities as blue circles with 1$\sigma$ error bars, and uncertainties of measured spectral indexes as bowties.  For any jet component non-detections, we show flux density upper limits as arrows.  Red arrows specifically denote \textit{Fermi}/LAT 95\% flux density upper limits which are key to assessing the viability of the IC/CMB model. Phenomenological synchrotron spectral models are fit by eye to the observed radio-to-optical flux densities using single power laws with high-frequency exponential cut-offs.  As described further in Appendix~\ref{images_SEDs}, the synchrotron spectral models have slopes chosen to best-match the observed X-ray spectral indexes (within the constraints of the radio flux densities and their associated errors), as this is a requirement of the IC/CMB model.

As described in \cite{georganopoulos2006} and \cite{meyer+17}, the IC/CMB spectral models have the same shape as the synchrotron spectral models but are shifted in frequency and luminosity by the amount necessary to match the observed X-ray flux densities.  We give the equations parametrizing these shifts in section~\ref{spectral models} (along with all of the other relevant details for our spectral models), but the pertinent feature of the IC/CMB model fits is that the IC/CMB spectral shape and associated predicted level of MeV-to-GeV gamma-ray flux is completely determined by the shape of the radio-to-optical synchrotron spectrum and the requirement of matching the observed X-ray flux densities.

In Figure~\ref{fig:sed_showcase}, the left-panel SED of knot~B from the jet hosted by the quasar PKS~1150+497 is a clear example of a case in which we can confidently rule out the IC/CMB model on the basis of its over-prediction of the large-scale-jet gamma-ray flux.  As is clearly shown in this SED, the red \textit{Fermi}/LAT flux density upper limits are well below the IC/CMB model predictions.  However, the right-panel SED for QSO~0957+561 in Figure~\ref{fig:sed_showcase} is a clear example where we can not rule out the IC/CMB model.  This is illustrated by the dotted synchrotron spectral model curve and corresponding dashed IC/CMB model curve which remains well below the \textit{Fermi}/LAT upper limits in the MeV-to-GeV bands.    

The gray circles in the SEDs of Figure~\ref{fig:sed_showcase} show where the observed radio-to-optical synchrotron flux densities would appear in the shifted IC/CMB spectra.  Since these portions of the IC/CMB spectra correspond to observed portions of the synchrotron spectra, there is no uncertainty in the IC/CMB model predictions in these regions of the SEDs.  Thus, we consider these gray points to be ``anchor points'' for the IC/CMB spectra. The case of QSO~0957+561 is an example of how uncertainty in the extrapolation from the observed GHz synchrotron flux densities to higher energies leads to uncertainty in the IC/CMB model predictions for the MeV-to-GeV spectrum.  If future high-frequency radio or optical/IR/UV observations are able to detect some portion of this undersampled synchrotron spectrum (especially the peak), this could change to a case in which the IC/CMB model is ruled out.

Applying the criteria for assessing the viability of the IC/CMB model described above, we find the IC/CMB model to be ruled out in 21/45 of the X-ray jets analyzed in this study on the basis of over-predicting the large-scale-jet MeV-to-GeV gamma-ray flux consistent with our \textit{Fermi}/LAT analysis.  Furthermore, we have ruled out the IC/CMB model in 18/36 of the jets from this sample which are unquestionably MSC X-ray jets and a single radio-to-X-ray synchrotron component is not viable.  In total, the IC/CMB model is now ruled out in 27/54 X-ray jets in which we have used the \textit{Fermi}/LAT to search for the required levels of gamma-ray flux under the IC/CMB model\footnote{This includes the addition of nine X-ray jets for which we have already published the \textit{Fermi}/LAT analysis \citep{meyer+14,meyer+15,breiding+17,meyer+19,roychowdhury+22}}.  It is important to note that we use the most conservative spectral models (given the data), with the lowest predicted MeV-to-GeV gamma-ray flux in assessing the viability of the IC/CMB model.  For the remainder of the paper, we will refer to the subset of sources for which the IC/CMB model is incompatible with the gamma-ray limits as the "ruled-out" sample, and the remainder as the "unconfirmed" sample.  When comparing the ``unconfirmed'' sample to the ``ruled out'' sample, it is important to stress that the ``unconfirmed'' sample may still contain X-ray jets which are not truly IC/CMB X-ray jets but for which we do not have sufficient data to rule out the IC/CMB hypothesis.  This has the potential to partially wash out sample differences, but it is still important to compare the two samples to assess any major correlations which still survive the potential sample contamination.

The particular sources from this study in which we have ruled out the IC/CMB model are designated in the last column of Table~\ref{tab:example_table}.  In Table~\ref{tab:example_table} we also give key properties describing our sample, including the source name, redshift, black hole mass, and angular-to-physical scale conversion factors.  Following the methodology outlined in \cite{meyer+11} and \cite{keenan+21} and using data obtained from the NASA/IPAC Extragalactic Database (NED), we also decomposed the radio spectra into the extended (isotropic) ``lobe'' components and beamed ``core'' components, allowing us to determine the radio core dominance, $\mathrm{R_{CE}}$. We define $\mathrm{R_{CE}}$ for our sample as the ratio of the modeled core-to-lobe luminosity at 1.4~GHz (in $\mathrm{\nu L_{\nu}}$), where these values can be found in Table~\ref{tab:example_table}.  Using these radio spectral decompositions, we also determined the isotropic lobe luminosities at 300~MHZ (in $\mathrm{\nu L_{\nu}}$), defined as $\mathrm{L_{ext}}$ .  We also estimate kinetic jet powers,  $\mathrm{L_{kin}}$, using the scaling relation from \cite{cavagnolo2010}.  Both $\mathrm{L_{ext}}$ and $\mathrm{L_{kin}}$ are also given in Table~\ref{tab:example_table}.

In Table~\ref{fermi-table} we give the minimum 95\% flux density upper limits (or minimum flux densities) obtained from our \textit{Fermi}/LAT data analysis, along with the predicted flux densities from the IC/CMB model and the upper limit on the Doppler factor, $\delta$, for the jet feature analyzed as a consequence of our \textit{Fermi}/LAT flux density upper limits (or minimum flux densities).  These $\delta$ limits are determined by the requirement that the IC/CMB spectra not violate the \textit{Fermi} flux density limits, and assume equipartition magnetic fields for the emission regions (given in Table~\ref{tab:example_table}, and measured\footnote{For these measurements we assume the EED $\mathrm{\gamma_{min}=10}$ and estimate radio jet volumes and flux densities from our images.} using the radio images presented in Figures~\ref{fig:img1}-\ref{fig:img2}).  The equipartition magnetic field strength assumes an equal energy density between particles and field, and is the minimum energy state for the system \citep{burbidge59}.

In Figures~\ref{fig:img1}-\ref{fig:img3} we show X-ray images with radio contours for each jet from this study in order to highlight the particular radio/X-ray morphologies for each jet and show the emission regions analyzed for our SEDs.  In Figures~\ref{fig:sed1}-\ref{fig:sed4} we show the jet SEDs complied with the multi-wavelength data described in section~\ref{methods}, in addition to the SED data taken from the literature, for each jet analyzed in this study.  We describe precisely where each flux density and spectral index measurement originates in section~\ref{source_notes}, where we also discuss other relevant properties of each source.

\section{Discussion}
\label{discussion}

As described in section~\ref{results}, we consider the IC/CMB model to be ruled out for a source if the 95\% \textit{Fermi} flux density upper limits (or minimum flux densities) are below the level required by the IC/CMB model (for some sources it is clearly ruled out at a greater confidence than 95\%).  For the sources in which the \textit{Fermi}/LAT flux density upper limits (or minimum flux densities) appear above the IC/CMB model prediction, we consider the IC/CMB model consistent with the \textit{Fermi}/LAT data, and classify the source as `unconfirmed'.  However, it is possible future \textit{Fermi}/LAT observations will yield deeper upper limits as integration time on-source accumulates, allowing for the possibility of ruling out the IC/CMB model in any of these jets in the future. Alternatively, it is possible enough accumulated on-source integration time may allow for us to detect the IC/CMB gamma-rays in the future from the unconfirmed sample jets, as demonstrated in \cite{meyer+19}.  The IC/CMB mechanism must still be operating at some level, even if it is not the primary emission process responsible for the anomalous X-rays observed in these jets.

\subsection{Alternative Models for the X-ray Emission}
\label{alternate_models}

The primary alternative model for the anomalously bright and/or hard X-ray production in MSC X-ray jets is synchrotron radiation originating from a second, higher-energy EED. If this is the correct physical description for any of the jets presented here, this implies highly efficient, \emph{in situ} particle acceleration many kpc from the central engine. These electrons would be much more energetic than the ones producing the observed radio-to-optical synchrotron spectrum (multi-TeV versus MeV-to-GeV energies).  This high-energy EED would in some cases need a low-energy cutoff such that it does not overproduce the observed, and often very faint or non-detected, optical/UV flux densities in many jets.  Furthermore, this distinct, high-energy EED will inverse-Compton scatter the CMB to TeV gamma-rays, potentially presenting a promising source class for observations with the upcoming Cherenkov Telescope Array \citep[CTA, ][]{meyer+15,georganopoulos2006}.  


\begin{figure}
	\includegraphics[scale=0.5]{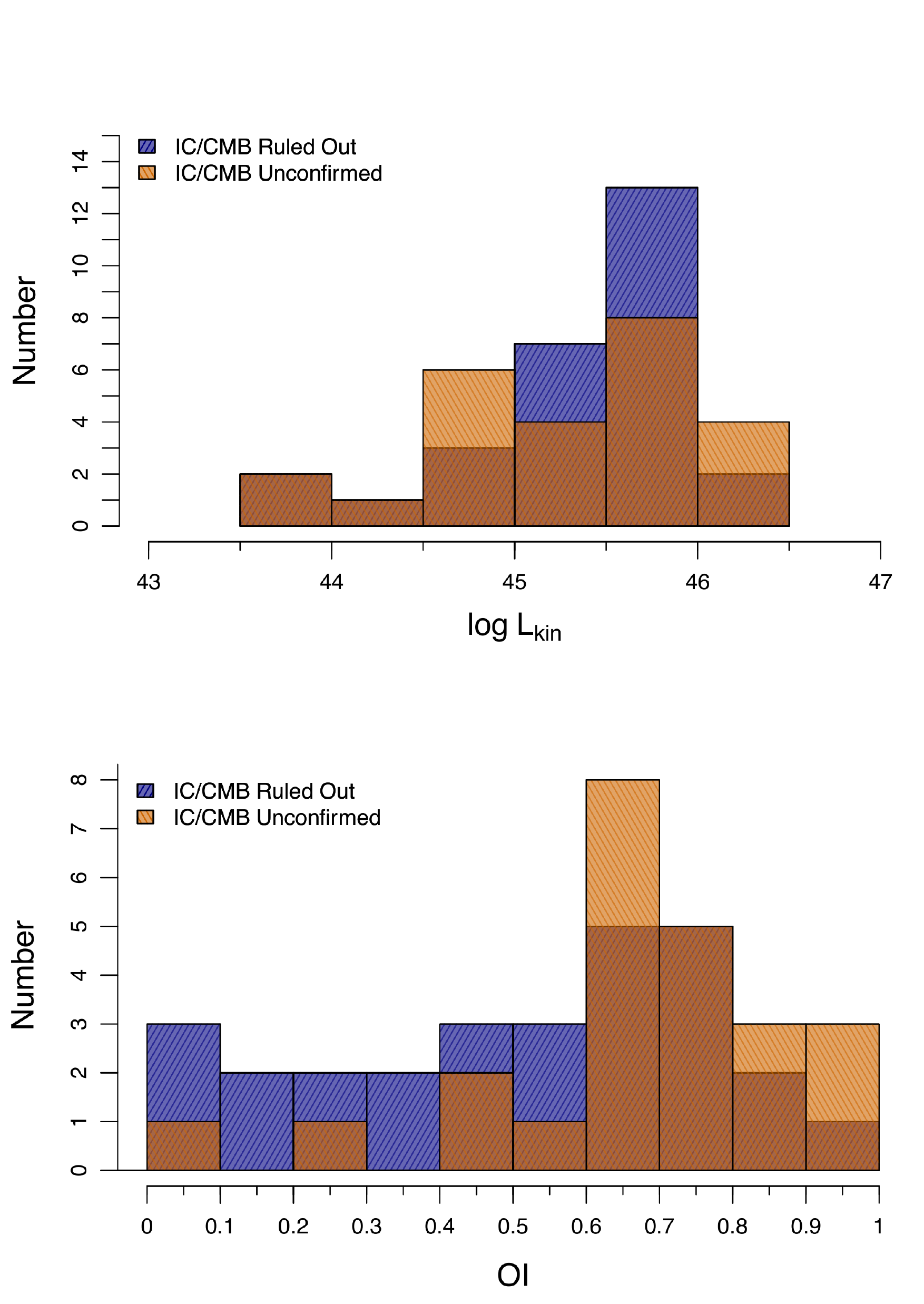}
	\caption{\textit{Top:} Histogram of kinetic jet powers, $\mathrm{L_{kin}}$, for the full sample of X-ray jets in which we have used \textit{Fermi} to search for IC/CMB gamma-rays.  The orange shaded region corresponds to the ``IC/CMB not ruled out jets'', and the purple shaded region corresponds to the ``IC/CMB ruled out jets''.  \textit{Bottom:} Histogram of the orientation indicator, OI, where 0 corresponds to misaligned jets and 1 the aligned jets.  The orange shaded region corresponds to the ``IC/CMB not ruled out jets'', and the purple shaded region corresponds to the ``IC/CMB ruled out jets''. }
	\label{fig:OI_LKIN}
\end{figure}
  
Alternatively, it is possible the anomalous X-rays originate via hadronic emission processes, where the X-rays result either from direct (relativistic) proton synchrotron radiation, or from the synchrotron radiation
produced by secondary electrons \citep[e.g.,][]{petropoulou2017,kusunose17}.  Hadronic models have a large number of free parameters, and thus it is difficult to test such models with e.g., SED-fitting.  However, one unique prediction from hadronic models is significant levels of neutrino flux \citep{mannheim89}, where spatio-temporal correlations with blazar gamma-ray flares observed by IceCube over the past few years offer some support for this scenario \citep[e.g.,][]{kadler}.  Additionally, similar to the IC/CMB model, hadronic models imply highly super-Eddington jets,
which offers challenges to the current accretion paradigm in AGN \citep[e.g.,][]{zdziarski+15}. However, \cite{tchekhovskoy+10} argue that highly super-Eddington jets might be accommodated if the jets extract a significant fraction of their energy from black hole spin via the Blandaford-Znajek mechanism \citep{blandford+77}.  

In theory, observations from a high-angular resolution X-ray polarimeter such as the \textit{Imaging X-ray Polarimetry Explorer} \citep[\textit{IXPE},][]{ixpe} could potentially provide another independant line of evidence against the IC/CMB model, since the X-rays resulting from the IC/CMB mechanism are predicted to have very low polarization (in contrast to the X-rays from both leptonic synchrotron and hadronic models). However, the relatively large PSF ($\sim$6.4~arcsec full width half maximum at 2~keV) of \textit{IXPE} will likely limit this technique to only the largest jets in angular size (e.g., the hotspot of Pictor A). Future observations with the Square Kilometer Array \citep[SKA,][]{ska} will have the requisite frequency coverage, sensitivity, and angular resolution to search for the very-low-frequency radio light (i.e., tens-to-hundreds of MHz) produced by the synchrotron radiation of the putative electrons directly up-scattering the IC/CMB X-rays (i.e., the low-energy extension of the GHz-emitting electrons with individual Lorentz factors of $\gamma \sim 100$ in the rest frame of the bulk-plasma flow).  Determining the correct physical model which accounts for the anomalously bright and/or hard X-ray production in MSC X-ray jets is crucial in determining the physical nature of these jets and the SMBHs which give rise to them, in addition to what role they play in interacting with their environments. 

For all of the X-ray jets in which the IC/CMB has now been ruled out by \textit{Fermi} observations, we can infer that these jets are not as closely aligned to the line of sight, are not as powerful, and are not as relativistic on kpc scales as would be required under the IC/CMB model.  Next we discuss some additional properties of the combined sample of 54 jets for which we have now applied the \textit{Fermi} test, and how they relate to the possible physical processes responsible for the observed X-ray emission.

\subsection{Jet Alignment \& Kinetic Powers}
\label{MISALIGNED}

One requirement of the IC/CMB model is very small jet angles to the line of sight in order to reproduce the observed, bright X-ray flux densities via relativistic beaming \citep{tavecchio2000,celotti2001}.  The radio core dominance, $\mathrm{R_{CE}}$, is typically used as a rough measure of relative jet misalignment in radio-loud AGN \citep[e.g.,][]{frederic+16}.  Qualitatively, this correlation between jet viewing angle and $\mathrm{R_{CE}}$ is expected to result from the fact that the extended lobe contribution is isotropic while the radio core emission is highly relativistically beamed.  Therefore, if any given jet were viewed at progressively larger angles to the observers line of sight, the core should show a commensurate drop in luminosity due to the radiation being increasingly beamed away from the observer while the lobe luminosity should remain fixed. 
However, it was shown in \cite{keenan+21} that the distribution of $\mathrm{R_{CE}}$ is dependant on kinetic jet power, where the highest-power jets at $\mathrm{log\left( L_{ext}\right)>43.5}$ have a maximum $\mathrm{log\left(R_{CE}\right)}$ of $\sim$1 ($\mathrm{L_{ext}}$ being a proxy for jet power).  Additionally, the peak of the distribution of $\mathrm{log\left(R_{CE}\right)}$ for low-power jets at $\mathrm{log\left( L_{ext}\right)<41.5}$ is $\sim0.5-1$ decades below that of high-power jets.  Therefore, we use the orientation indicator, OI, introduced in \cite{keenan+21} as our measure of relative jet misalignment in our sample of jets, which accounts for this discrepancy in jet power by also including the core luminosity in the parametrization.  The equation for the OI is

\begin{equation}
	\mathrm{OI=0.077~log\left(L_{core}\right) + 0.021~log\left(R_{CE}\right)-2.615}
\end{equation}         

where $\mathrm{L_{core}}$ is the  $\mathrm{\nu L_{\nu}}$ core luminosity.  Values of 0 correspond to the most misaligned jets (i.e., radio galaxies with jets aligned in the plane of the sky) and 1 the most aligned (i.e., highly aligned blazars viewed jet-on), where we truncate any values below or above these boundaries as described in \cite{keenan+21}.  

In Figure~\ref{fig:OI_LKIN} we show histograms of the kinetic jet power, $\mathrm{L_{kin}}$, and OI, for our full sample of 54 jets.  We discern no significant difference in the ruled-out versus unconfirmed jets in terms of $\mathrm{L_{kin}}$.  The IC/CMB model generally predicts more powerful jets, so this is already at odds with the idea that the unconfirmed sources are dominated by IC/CMB X-ray jets.  Interestingly, we do find support for the unconfirmed sources showing some skew towards more aligned jets than the ruled-out jets when comparing the OI histograms.  This may be indicating that IC/CMB X-ray jets are hidden in our unconfirmed sample, as a high degree of jet alignment is a key feature of the IC/CMB model.  Alternatively, it may be the case that more aligned jets are brighter \textit{Fermi}/LAT gamma-ray sources.  Thus, the \textit{Fermi}/LAT observations will be less constraining in the more aligned jets than the misaligned jets, making it harder to rule out the IC/CMB model in the aligned jets.  This scenario is partially supported by the fact the \textit{Fermi}/LAT 4FGL source catalog is overwhelmingly dominated by blazars \citep[][although there are certainly some nearby misaligned jets in the catalog]{4fgl}.  Future \textit{Fermi}/LAT observations will help address which of these scenarios is correct.  As \emph{Fermi} continues to monitor the whole sky, it is likely that more sources will either join the ranks of the ruled-out sample, or finally show the expected plateau signature from the large-scale jet.

\begin{figure}
	\includegraphics[scale=0.5]{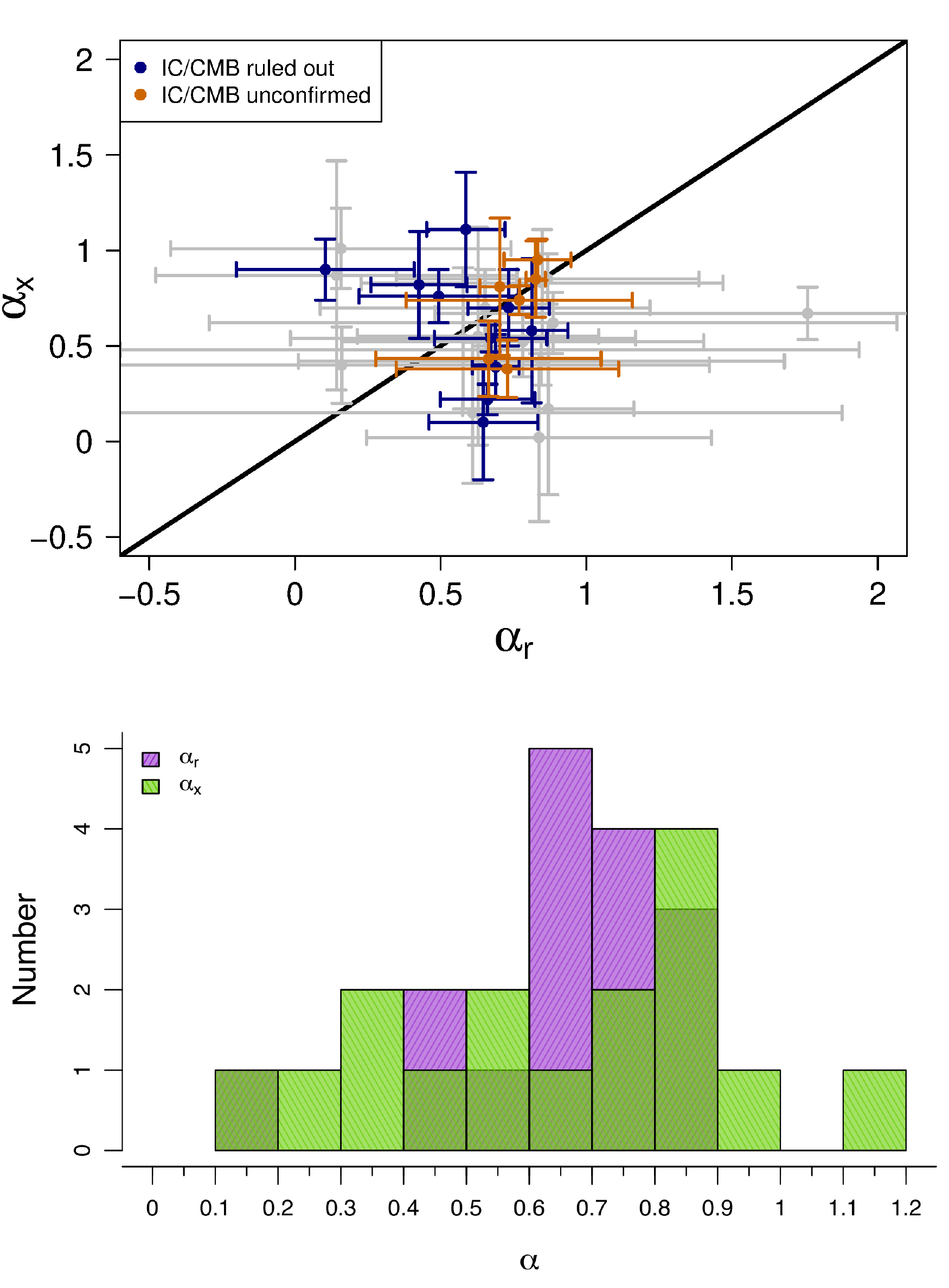}
	\caption{\textit{Top:} The X-ray spectral index plotted against the radio spectral index for each jet analyzed in this study.  The jet regions analyzed are given in Table~\ref{fermi-table} and denoted in the SEDs in Figures~\ref{fig:sed1}-\ref{fig:sed4}.  We plot in gray the least-well-measured indexes, where at least the radio or X-ray spectral index error exceeds 0.4.  The colored indexes correspond to those in which both the radio and X-ray spectral index errors are less than 0.4.  The orange indexes corresponds to the ``IC/CMB not ruled out jets'', and the purple indexes correspond to the ``IC/CMB ruled out jets''.  \textit{Bottom:} Histograms of the X-ray and radio spectral indexes ($\mathrm{\alpha_{x}}$ and $\mathrm{\alpha_{r}}$, respectively) for all of the jet regions analyzed in this study ---  $\mathrm{\alpha_{r}}$ is shown in purple and $\mathrm{\alpha_{x}}$ in green.}
	\label{fig:indexes}
\end{figure}

\begin{table}
		\caption{Spectral Index Correlation Tests} 
		\label{index-table}
		\begin{tabular}{l c c c c c}
			\hline
			\centering
			Correlation&&&P-Value&&\% Tests\\
			Subset&$\mathrm{\bar{\rho_{s}}}$&$\mathrm{\sigma_{s}}$ &68\% Conf.&N&Failed\\
			\hline
			All&-0.07&0.17&[0.28,0.91]&32&98\\\\
			All Well-Measured&-0.05&0.25&[0.23,0.87]&16&97\\\\
			Well-Measured &-0.26&0.30&[0.45,0.96]&6&99\\
			ruled-out&&&&&\\\\
			Well-Measured &0.13&0.44&[0.12,0.73]&10&97\\
			unconfirmed&&&&&\\
			\hline \\
		\end{tabular}
\end{table}

\subsection{X-ray \& Radio Spectral Indexes}
\label{indexes}

In Figure~\ref{fig:indexes}, we show the X-ray and radio spectral index ($\mathrm{\alpha_{x}}$ and $\mathrm{\alpha_{r}}$, respectively) histograms for the X-ray jet regions analyzed in the SEDs for this study.  
In this Figure we also plot the radio spectral index against the X-ray spectral index for each jet region analyzed.  The expectation from the IC/CMB model is that the radio spectral index should be the same as the X-ray spectral index, assuming the observed GHz-frequency radio spectral index is the same as the much lower and unobserved tens-to-hundreds of MHz-frequency spectral index probing the portion of the EED directly responsible for the IC/CMB X-rays.  In gray we plot the indexes with errors $>0.4$, where the colored indexes have both radio and X-ray spectral index errors $<0.4$ and we consider these the well-measured indexes.  
In purple are the well-measured indexes for the ruled-out cases and orange denote the IC/CMB unconfirmed subsample.  

While it does not appear that $\mathrm{\alpha_{x}}$ tends to be either harder or softer than $\mathrm{\alpha_{r}}$, the question remains whether there is any degree of correlation.   
The simplest form of the IC/CMB model predicts a linear correlation between $\mathrm{\alpha_{r}}$ and $\mathrm{\alpha_{x}}$. To test this we used the non-parametric Spearman's rank-order correlation coefficient \citep{spearman1904}, $\mathrm{\rho_{s}}$, which assesses the presence of any monotonic dependence of one variable on another and is in general a more robust indicator of correlation than the Pearson correlation coefficient which assesses the degree of linear correlation and can be overly sensitive to outliers \citep[e.g., ][]{wilcox04}.

We also looked for any positive dependence between the two quantities at all.  $\mathrm{\rho_{s}}$ values of +1 correspond to high-degrees of positive correlation, 0 corresponds to uncorrelated data, and -1 to highly negatively correlated data.  We determined the significance for $\mathrm{\rho_{s}}$ using Student's t-test \citep{zar72}.  Following the methodology outlined in \cite{curran14}, we determine empirical distributions of $\mathrm{\rho_{s}}$ by performing 10,000 trials of bootsrapping with perturbation on the ($\mathrm{\alpha_{r}}$, $\mathrm{\alpha_{x}}$) data pairs.  Similar to conventional bootstrapping, this methodology relies on Monte Carlo random sampling with replacement of each ($\mathrm{\alpha_{r}}$, $\mathrm{\alpha_{x}}$) data pair, but also adds some perturbation to the value of each ($\mathrm{\alpha_{r}}$, $\mathrm{\alpha_{x}}$).  In this case, we used a perturbative term randomly chosen from a normal distribution with a mean of zero and standard deviation corresponding to our index error, but this method does not in general rely on normally distributed errors for the data.  We ran these bootstrap trials for all indexes, all well-measured indexes, the well-measured indexes belonging to the ruled-out subsample, and the well-measured indexes belonging to the unconfirmed subsample.  

\begin{table}
	\caption{$\mathrm{log\left(R_{x}\right)}$ vs $\mathrm{log\left(R_{CE}\right)}$ Correlation Tests} 
	\label{xdom-table}
	\begin{tabular}{l c c c c c}
		\hline
		\centering
		Correlation&&&P-Value&&\% Tests\\
		Subset&$\mathrm{\bar{\rho_{s}}}$&$\mathrm{\sigma_{s}}$ &68\% Conf.&N&Failed\\
		\hline
		All&0.39&0.11&[0.00007,0.02]&54&8\\
		 ruled-out&0.36&0.16&[0.002,0.15]&28&36\\
		unconfirmed&0.45&0.16&[0.0006,0.08]&26&22\\
		\hline \\
	\end{tabular}
\end{table} 

\begin{table}
	\caption{Model Parameters for Misalignment Curves in Figure~\ref{fig:xdom}} 
	\label{model-params}
	\begin{tabular}{l c c c c c c c }
		\hline
		\centering
		Curve&&X-ray&Radio&&&&\\
		Name&Panel&Zone&Zone&$\mathrm{\Gamma_{C}}$&$\mathrm{\Gamma_{X}}$&$\mathrm{\Gamma_{R}}$&$\mathrm{\alpha_{X}}$\\
		\hline
		IC/CMB&I&...&...&10&10&10&0.7\\
		&&&&100&10&10&0.7\\
		2nd Synch.&II&Sheath&Spine&10&1.5&2.5&0.7\\
		&&&&10&2&2.5&0.7\\
		2nd Synch.&III&Spine&Sheath&10&2.5&1.5&0.7\\
		&&&&10&1.9&1.5&0.7\\
		Two-Zone&IV&Spine&Sheath&10&3.5&2.5&0\\
		IC/CMB&&&&10&1.9&1.5&0\\
		\hline \\
	\end{tabular}
\end{table}

\begin{figure*}
	\includegraphics[scale=0.5]{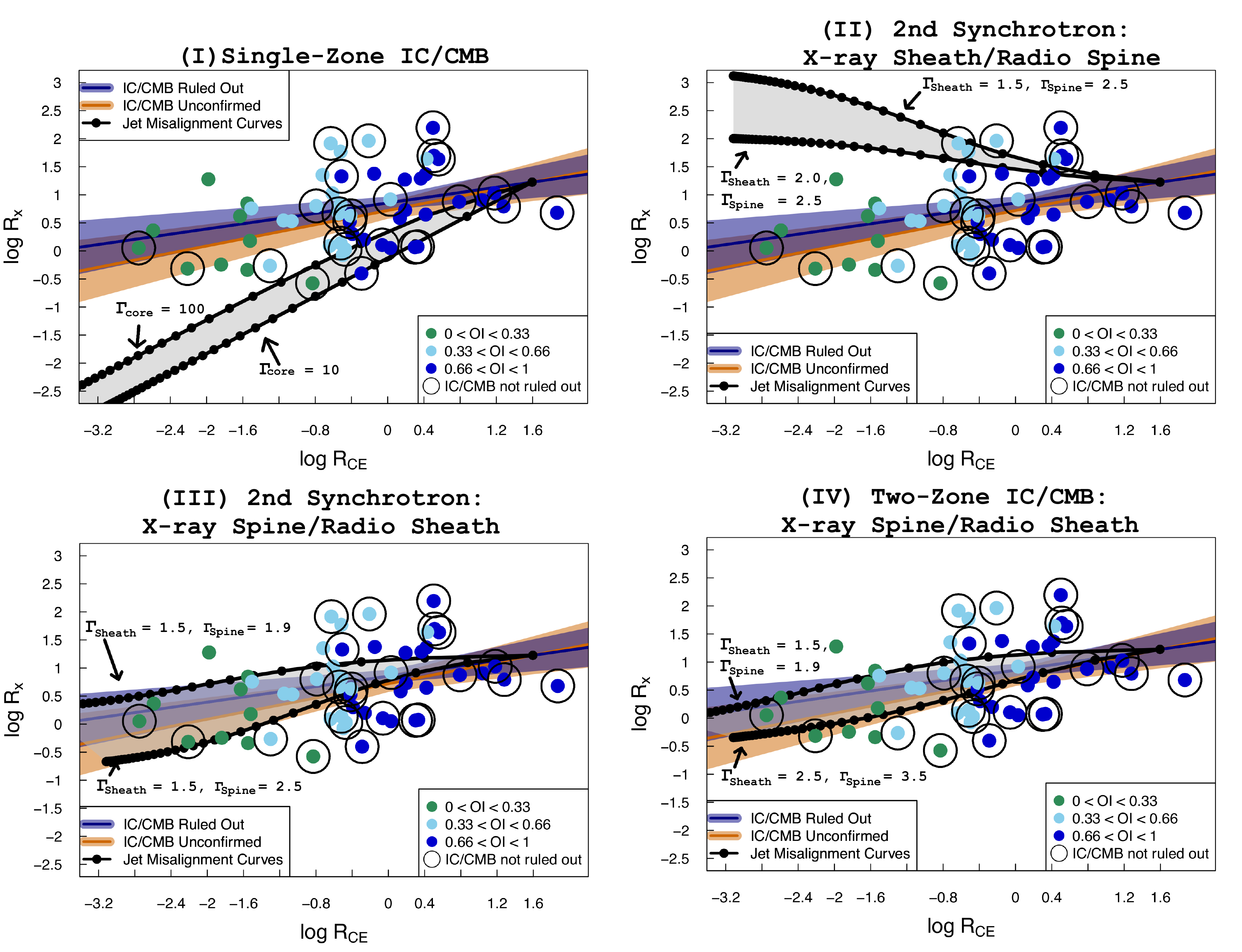}
	\caption{The X-ray dominance, $\mathrm{R_{x}}$, for the jet features analyzed in Figures~\ref{fig:sed1}-\ref{fig:sed4}, plotted against the core dominance, $\mathrm{R_{CE}}$, for our full sample of 54 jets we have now used \textit{Fermi}/LAT observations to test the IC/CMB model (in log-log space).  We binned the jets according to their alignment to the line of sight, assigning different colors to each bin: $\mathrm{0<OI<0.33}$ (most misaligned, green), $\mathrm{0.33<OI<0.66}$ (medium alignment, sky blue), and $\mathrm{0.66<OI<1}$ (most aligned, royal blue).  As expected, $\mathrm{R_{CE}}$ reasonably tracks jet orientation as parametrized by OI.  We also placed black circles around the jets in which we have ruled out the IC/CMB model with \textit{Fermi}/LAT results.  Error bars are not shown as they are on the order of the size of the plotted points.  We plot in each panel the mean OLS linear regression lines (plus 68\% confidence regions) for the ruled-out and unconfirmed subsamples, in purple and orange, respectively.  Each different panel corresponds to the misalignment of a fiducial jet under different jet models as described in section~\ref{BEAMING PATTERNS}, where the data for each panel is the same.  The jet parameters describing each black misalignment curve are given in Table~\ref{xdom-table}.  The black circles on the misalignment curves mark incremental 5 degree misalignments to the line of sight, starting at an initial alignment of five degrees.  The gray zones between the misalignment curves show the expected misalignment for parameter values varying between those given in Table~\ref{xdom-table}.}
	\label{fig:xdom}
\end{figure*}

In Table~\ref{index-table} we give the results of these correlation tests, including the mean and standard deviation of $\mathrm{\rho_{s}}$ ($\mathrm{\bar{\rho_{s}}}$ and $\mathrm{\sigma_{s}}$, respectively), 68\% confidence limits for the corresponding p-values, the number of data pairs for each correlation test (N), and percent of bootstrapped trials showing no degree of positive correlation.  In this case, failing the significance test is defined as having a p-value $>0.05$ for a one-tailed Student's t-test, meaning we can not reject the null hypothesis at $>95\%$ confidence that $\mathrm{\alpha_{r}}$ is uncorrelated with $\mathrm{\alpha_{x}}$.  Interestingly, the correlation tests between $\mathrm{\alpha_{r}}$ and $\mathrm{\alpha_{x}}$ for all subsets considered overwhelmingly fail to find a significant correlation.  This result suggests that distinct particle populations are responsible for the observed radio and X-ray emission in these jets, arising from either different physical processes, different physical conditions, or some combination thereof.  This is clearly in conflict with the expectations from the IC/CMB model, but a larger sample size of precise spectral index measurements would be desirable for confirming this result.

\subsection{X-ray \& Radio Beaming Patterns}
\label{BEAMING PATTERNS}


X-rays from the IC/CMB mechanism are expected to be highly relativstically beamed, meaning highly focused along the axis of the bulk-plasma flow in these jets.  The equation governing the relativistic beaming pattern for IC/CMB radiation is 

\begin{equation}
	\label{iccmb_beaming}
	\mathrm{L=L'\delta^{p+1+2\alpha_{r}}}
\end{equation}   

\noindent
with p=2 for a continuous flow and p=3 for discrete moving blobs \citep{dermer95,georganopoulos01}.  Here, L is the luminosity in the galaxy rest frame assuming isotropy, $\mathrm{L'}$ is the solid-angle-integrated luminosity in the jet frame, $\mathrm{\alpha_{r}}$ is the radio spectral index, and $\delta$ is the Doppler factor.  In the case of synchrotron emission, the expected beaming pattern is given by \citep{dermer95}
 
\begin{equation}
 	\label{synch_beaming}
 	\mathrm{L=L'\delta^{p+\alpha_{r}}}.
\end{equation}

\noindent
Increasing the angle of the jet to the line of sight decreases the value of $\delta$. Therefore, the observed IC/CMB X-ray flux should decrease faster than the synchrotron radio flux if viewed at progressively larger angles to the line of sight due to the stronger dependence on $\delta$ for IC/CMB.  We define a parameter known as the X-ray dominance, $\mathrm{R_{x}}$, which is the ratio of the 1~keV X-ray flux density to the 8.6~GHz radio flux density for the extended jet regions analyzed in our SEDs (in $\mathrm{\nu F_{\nu}}$, where the values of $\mathrm{R_{x}}$ for our sample can be found in Table~\ref{tab:example_table}).  In the four panels of Figure~\ref{fig:xdom}, we show log-space plots of $\mathrm{R_{x}}$ against $\mathrm{R_{CE}}$ for the full sample of 54 X-ray-jet features analyzed in this and other recent publications.  We binned the jets into three different groups corresponding to their relative alignment to the line of sight, assigning different colors to each bin: $\mathrm{0<OI<0.33}$ (most misaligned, green), $\mathrm{0.33<OI<0.66}$ (medium alignment, sky blue), and $\mathrm{0.66<OI<1}$ (most aligned, royal blue).  We also placed black circles around the jets in which we have ruled out the IC/CMB model with \textit{Fermi}/LAT results in order to compare any potential correlations  between the ruled-out and the unconfirmed groups.  Following the same methodology as in section~\ref{indexes}, we tested for any monotonic correlations between $\mathrm{log\left(R_{x}\right)}$ and $\mathrm{log\left(R_{CE}\right)}$ by computing empirical distributions of $\mathrm{\rho_{s}}$ based upon 10,000 boostrapped trials (in this case we did not add perturbation to the bootstrapping since errors on $\mathrm{log~R_{x}}$ and $\mathrm{log~R_{CE}}$ are negligible).  We performed correlation tests for subsets consisting of all of the jets, the ruled-out subsample, and the unconfirmed subsample; the results of these correlation tests are given in Table~\ref{xdom-table}.  

The majority of bootstrap trials in all of these subsamples, and mean $\mathrm{\rho_{s}}$ values for each subsample, indicated significant (p-value$<0.05$) positive correlations. Not surprisingly the most significant results were for the full sample of jets, likely due to the larger sample size.  Accordingly, and since we have no expectation for the functional form of the monotonic dependence of $\mathrm{log\left(R_{x}\right)}$ on $\mathrm{log\left(R_{CE}\right)}$, we fit ordinary least squares (OLS) linear regression lines to both the ruled-out and unconfirmed subsamples in the 10,000 bootstrapped trials.  We plot in all panels of Figure~\ref{fig:xdom} the resulting mean best-fit regression lines with corresponding shaded 68\% confidence intervals constructed from the bootstrapped statistics.  As is evident, the 68\% confidence regions of the OLS regression lines are largely overlapping, indicating no significant difference in the best-fit slope between the ruled-out and unconfirmed groups.  

To demonstrate in Figure~\ref{fig:xdom} how a single hypothetical jet would move through the plane when misaligned, we plot as connected black points the  `de-beaming path' for different assumed models of X-ray emission, using equation~\ref{iccmb_beaming} (for IC/CMB) and equation~\ref{synch_beaming} (for synchrotron).  This fiducial jet is assumed to be at an initial alignment of 5 degrees to the line of sight and connected points are spaced by 5 degrees.  We show two curves corresponding to different jet model parameters for each panel of Figure~\ref{fig:xdom}, as given in Table~\ref{model-params}.  We shown in gray the expected misalignment zone corresponding to a range of jet model parameter values between those which delineate the two curves.  These misalignment curves assume a flat radio core spectral index of 0, a radio spectral index for the observed large-scale-jet radio emission of 0.7, and a value of p equal to 0 for the core and 1 for the large-scale-jet component.  Additionally, this analysis assumes the angle of the (kpc-scale) large-scale-jet to the line of sight to be the same as that for the pc-scale core.

In panel I of Figure~\ref{fig:xdom} we show the expected misalignment of the fiducial jet under the standard single zone IC/CMB model we have been testing in this study\footnote{Using equations~\ref{iccmb_beaming}~\&~\ref{synch_beaming}, it can be shown that $\mathrm{R_{CE}\propto \delta_{core}^{p+\alpha_{core}}}$ and $\mathrm{R_{x}\propto \delta_{knot}^{1+\alpha_{knot}}}$ under the IC/CMB model.  Here, $\mathrm{\delta_{knot}}$ is the Doppler factor of the large-scale-jet component, $\mathrm{\delta_{core}}$ is the Doppler factor of the core, $\mathrm{\alpha_{knot}}$ is the radio spectral index for the large-scale-jet component, and $\mathrm{\alpha_{core}}$ is the core spectral index.}.   For this case, the misalignment zone is entirely determined by a range of core bulk-flow Lorentz factors, $\mathrm{\Gamma_{C}}$, between 10 and 100.  It is clear that the observed correlation between $\mathrm{log\left(R_{x}\right)}$ and $\mathrm{log\left(R_{CE}\right)}$ is much less steep than the IC/CMB model predictions --- albeit with a large degree of scatter which can be accounted for by individual jet-to-jet variations.  This suggests that the X-rays in these jet regions are only moderately stronger beamed than the radio synchrotron radiation.  While we can not rule out the possibility that a few (single-zone) IC/CMB X-ray jets are hidden in Figure~\ref{fig:xdom}, they are manifestly not the dominant type of X-ray jet in our sample.  Furthermore, the existence of a correlation for the ruled-out subsample already necessitates another physical model, and the similarity in regression slope to the unconfirmed subsample suggests they may be of similar physical origin.  However, larger sample sizes are needed to confirm these correlations and help tease apart the physical mechanism(s) responsible.  Below we discuss different multi-zone emission models which can account for the observed $\mathrm{log\left(R_{x}\right)}-\mathrm{log\left(R_{CE}\right)}$ correlation through the adjustment of an increased number of tuneable model parameters.

In panels II-IV of Figure~\ref{fig:xdom}, we consider two-zone models for the large-scale jet in which there is a fast-flowing jet spine surrounded by a slower-moving jet sheath \citep[e.g.,][]{sol+89,celotti2001}.  In all of these two-zone models, we only consider ``mildly relativstic'' large-scale jets with $\Gamma$ ranging between 1.5 and 3.5, and more relativistic cores with $\mathrm{\Gamma_{C}}=10$.  We differentiate the Lorentz factors between the X-ray-emitting zones and radio-synchrotron emitting zones by $\mathrm{\Gamma_{X}}$ and $\mathrm{\Gamma_{R}}$, respectively, which are given in Table~\ref{xdom-table}.  We also give the spectral index for the particles responsible for the X-rays in Table~\ref{xdom-table}, $\mathrm{\alpha_{X}}$.  In the context of 2nd synchrotron models, $\mathrm{\alpha_{X}}$ corresponds to the observed X-ray spectral index for the jet region which is assumed to be synchrotron in origin.  In the standard IC/CMB or ``two-zone'' IC/CMB models, $\mathrm{\alpha_{X}}$ corresponds to the radio spectral index of the low-energy electrons upscattering the CMB to X-ray energies.  For the standard IC/CMB model, we assume $\mathrm{\alpha_{X}}$ is the same as the observed GHz radio spectral index for the large-scale jets.  However, for the ``two-zone'' IC/CMB model in panel IV, $\mathrm{\alpha_{X}}$ corresponds to an electron population in the jet spine while the observed radio synchrotron emanates from the jet sheath.   

As is evident from Figure~\ref{fig:xdom}, both the 2nd synchrotron and ``two-zone'' IC/CMB models of panels III and IV can account for the more mildly beamed X-rays in a manner which is consistent with the observed correlation.  What these models have in common is that the X-rays originate from particles in a faster-moving spine, while the radio synchrotron emission originates from particles in a slower-moving surrounding sheath layer.  In panel II, we show a two-zone 2nd synchrotron model in which the X-rays originate from the slower-moving sheath and the radio originates from a faster-moving spine.  In this case, we would expect a negative correlation in the $\mathrm{log\left(R_{x}\right)}-\mathrm{log\left(R_{CE}\right)}$ plane, as opposed to the positive correlation we observe.

These results imply that these X-ray jet regions should be narrower than the corresponding radio-jet regions in these systems, and should be resolvable for very nearby jets.  In fact, X-ray jets which are narrower than their radio counterparts have been observed in several cases, one noteable example being the MSC X-ray jet in 3C~353 \citep{kataoka+08}.  Interestingly, the X-ray emission in 3C~353 is modeled as a 2nd synchrotron component by the authors, where the IC/CMB model is ruled out on the basis of an X-ray counter-jet detection.  Also of note in the context of 2nd synchrotron models is the observation of optical jets which are narrower than their radio-jet counterparts \citep[e.g.,][]{sparks+94}.  Together, these results suggest that if we are observing a 2nd, higher-energy synchrotron component in these jets, the high-energy particles are accelerated in the faster-moving inner spine and not the sheath boundary layers as suggested in shear-layer acceleration models \citep[e.g.,][]{tavechhio+21}.  However, as shown in panel IV, we can not rule out the possibility that a ``two zone'' IC/CMB model is operating in some of these jets.  This is different from the (single-zone) IC/CMB model we have been testing in this paper which requires the same electron population that is emitting the radio-to-optical synchrotron radiation to be responsible for up-scattering the CMB to X-rays.  In the type of two-zone IC/CMB model shown in panel IV, the electrons up-scattering the CMB to X-rays are in the spine and distinct from the electron population in the sheath responsible for producing the observed radio synchrotron spectrum.  However, even though this type of two-zone IC/CMB model is more consistent with the observed $\mathrm{log\left(R_{x}\right)}-\mathrm{log\left(R_{CE}\right)}$ correlation and has not been ruled out by our \textit{Fermi} test, it suffers from many of the same problems as the single-zone IC/CMB model (e.g., knotty X-ray jets, X-ray flux variability, etc.) and may only be applicable to the most aligned jets which remain highly relativistic on kpc scales.


\subsection{Evidence For Jet Deceleration}
\label{DECEL}

The IC/CMB model requires large bulk-flow Lorentz factors, $\Gamma$, on kpc scales (i.e. highly relativistic, $\Gamma\sim10$) in addition to being oriented at small angles to the line of sight.  Very Long Baseline Interferometry (VLBI) observations of apparent superluminal motion for features at the base of AGN jets provide direct evidence for highly relativistic bulk-flow speeds on pc scales \citep[e.g.,][]{lister+13}. Measurement of the maximum apparent speed ($\mathrm{\beta_{app}}$, the apparent speed of the jet feature in units of the speed of light, c) for these features, gives the most stringent lower limit on $\Gamma$ and upper limit on the angle of the jet to the line of sight on the pc scale. The \textit{Fermi} limits presented in this work allow us to place upper limits on $\mathrm{\delta/B}$ for the kpc-scale jet such that the IC/CMB model curves fall below these limits. If we further assume that the magnetic field is at its equipartition value (these values are given in Table~\ref{tab:example_table}), then these observations give limits on the Doppler factor, $\delta$ (these limits can be found in Table~\ref{fermi-table}).  

In Table~\ref{decel-table}, we show the maximum $\mathrm{\beta_{app}}$ values obtained with VLBI (which we label as $\mathrm{\beta_{max}}$) for the sources in which we could find measurements in the literature, as noted in Table~\ref{decel-table}. In Table~\ref{decel-table} we also give the maximum angle of the jet to the line of sight, $\mathrm{\theta_{max}}$, and minimum bulk-flow Lorentz factor, $\mathrm{\Gamma_{min,~pc}}$, which are consistent with the measured values of $\mathrm{\beta_{max}}$\footnote{It is a straightforward exercise to show that $\mathrm{\Gamma_{min,~pc}=\frac{\beta_{max}}{\beta}\approxeq\beta_{max}}$ (where $\beta$ is the true physical speed of the jet flow in units of c) and $\mathrm{\theta < cos^{-1}\left(\frac{\beta_{max}^{2}-1}{\beta_{max}^{2}+1}\right)\equiv\theta_{max}}$.} (where $\mathrm{\theta_{max}}$ and
$\mathrm{\Gamma_{min,~pc}}$ apply to the pc-scale portion of the base of the jet). Finally, in
the last column of Table~\ref{decel-table} we give the upper limit to the bulk-flow Lorentz factor, $\mathrm{\Gamma_{max,~kpc}}$, for the kpc-scale jet feature analyzed in our jet SEDs. $\mathrm{\Gamma_{max,~kpc}}$ is determined by our upper
limits on the Doppler factor, assuming an equipartition magnetic field. We additionally give error bars on $\mathrm{\Gamma_{max,~kpc}}$ representing the $\mathrm{\Gamma_{max,~kpc}}$ limits corresponding to a magnetic field strength a factor of two out of equipartition. Furthermore, we assume the jet angle to the line of sight for the kpc-scale jet is less than $\mathrm{\theta_{max}}$. This assumption is
reasonable allowing that the jet does not significantly bend in the plane of observation.  

\begin{table}
	\begin{threeparttable}
		\caption{Jet Deceleration Parameters}
		\label{decel-table} 
		\begin{tabular}{l l c c r}
			\hline
			\centering
			Source&Common&&&\\
			Name & Name&$\mathrm{\beta_{max}}$&$\mathrm{\theta_{max}}$&$\mathrm{\Gamma_{max,~kpc}}$\\
			(J2000)&&&($^{\circ}$)&\\ 
			\hline
			J0108+0135\tnote{$\dagger$}&PKS~B0106+013&26.8$\mathrm{^{c}}$&4.27&4.35$^{+\ 2.77}_{-\ 1.40}$\\[5pt]
			J0237+2848&4C~28.07&27.3$\mathrm{^{e}}$&4.20&8.41$^{+\ }_{-\ 3.53}$\\[5pt]
			J0418+3801&3C~111&8.42$\mathrm{^{a}}$&13.5&...\\[5pt]
			J0433+0521&3C~120&9.09$\mathrm{^{f}}$&12.6&...\\[5pt]
			J0607$-$0834\tnote{$\dagger$}&PKS~0605$-$085&32.8$\mathrm{^{a}}$&3.49&6.42$^{+\ 5.98}_{-\ 2.16}$\\[5pt]
			J0728+6748\tnote{$\dagger$}&3C~179&8.99$\mathrm{^{i}}$&12.7&1.54$^{+\ 0.768}_{-\ 0.350}$\\[5pt]
			J0741+3112&B2~0738+31&11.1$\mathrm{^{e}}$&10.3&...\\[5pt]
			J0830+2410&B2~0827+24&24.7$\mathrm{^{k}}$&4.64&5.74$^{+\ }_{-\ 2.08}$\\[5pt]
			J0840+1312\tnote{$\dagger$}&4C~13.38&13.3$\mathrm{^{e}}$&8.60&2.27$^{+\ 1.28}_{-\ 0.659}$\\[5pt]
			J0922$-$3959\tnote{$\dagger$}&PKS~0920$-$397&30.8$\mathrm{^{d}}$&3.72&2.23$^{+\ 0.893}_{-\ 0.584}$\\[5pt]
			J1048$-$1909&PKS~1045$-$188&10.9$\mathrm{^{b}}$&10.5&...\\[5pt]
			J1058+1951\tnote{$\dagger$}&4C~20.24&10.5$\mathrm{^{l}}$&10.9&1.48$^{+\ 0.589}_{-\ 0.315}$\\[5pt]
			J1130$-$1449\tnote{$\dagger$}&PKS~1127-145&31.6$\mathrm{^{k}}$&3.63&2.33$^{+\ 0.937}_{-\ 0.619}$\\[5pt]
			J1153+4931\tnote{$\dagger$}&PKS~1150+497&18.2$\mathrm{^{a}}$&6.29&3.71$^{+\ 3.24}_{-\ 1.27}$\\[5pt]
			J1205$-$2634&PKS~1202$-$262&11.0$\mathrm{^{a}}$&10.4&...\\[5pt]
			J1224+2122&4C~21.35&28.0$\mathrm{^{a}}$&4.09&9.98$^{+\ }_{-\ 4.47}$\\[5pt]
			J1632+8232&NGC~6251&0.156$\mathrm{^{b}}$&162&...\\[5pt]
			J1642+3948&3C~345&24.6$\mathrm{^{h}}$&4.66&...\\[5pt]
			J1642+6856\tnote{$\dagger$}&4C~69.21&24.9$\mathrm{^{c}}$&4.60&2.17$^{+\ 0.864}_{-\ 0.557}$\\[5pt]
			J1746+6226&4C~62.29&15.2$\mathrm{^{c}}$&7.53&...\\[5pt]
			J1829+4844&3C~380&15.4$\mathrm{^{l}}$&7.43&4.46$^{+\ }_{-\ 1.76}$\\[5pt]
			J1849+6705&8C~1849+670&31.6$\mathrm{^{e}}$&3.63&...\\[5pt]
			J1927+7358&4C~73.18&22.4$\mathrm{^{d}}$&5.11&...\\[5pt]
			J2005+7752&S5~2007+777&4.40$\mathrm{^{g}}$&25.6&...\\[5pt]
			J2158$-$1501&PKS~2155$-$152&21.5$\mathrm{^{j}}$&5.33&...\\[5pt]
			J2203+3145&4C~31.63&8.81$\mathrm{^{j}}$&13.0&...\\[5pt]
			J2218$-$0335&PKS~2216$-$038&6.91$\mathrm{^{a}}$&16.5&...\\[5pt]
			J2253+1608&3C~454.3&25.4$\mathrm{^{i}}$&4.51&..\\
			\hline \\
		\end{tabular}
		\begin{tablenotes}
			\item $\mathrm{\beta_{max}}$ measurements are obtained from the following literature sources (corresponding to the superscripts in the $\mathrm{\beta_{max}}$ column): (a) \cite{frey+15}, (b) \cite{homan+01}, (c) \cite{jiang+02}, (d) \cite{jorstad+05}, (e) \cite{jorstad+17}, (f) \cite{kellerman+04}, (g) \cite{lister+09}, (h) \cite{lister+13}, (i) \cite{lu+12}, (j) \cite{piner+01}, (k) \cite{piner+12}, (l) \cite{sudouandiguchi11}.
			\item[$\dagger$]These sources are the ones which make the plot of $\mathrm{\Gamma_{max,~kpc}}$ vs $\mathrm{\Gamma_{min,~pc}}$ shown in Figure~\ref{fig:decel}, where they all show  evidence of pc-to-kpc scale jet deceleration.  The sources in which an ellipses is given for $\mathrm{\Gamma_{max,~kpc}}$ have Doppler factor limits not constraining enough to be translated into Lortentz factor limits for the given $\mathrm{\beta_{max}}$ measurements.  The errors on $\mathrm{\Gamma_{max,~kpc}}$ represent magnetic field strengths a factor of two out of equipartition.  For cases in which the upper bound on $\mathrm{\Gamma_{max,~kpc}}$ could not be determined due to the implied $\delta$ limit for this magnetic field value, we leave this error value empty.   
		\end{tablenotes}
	\end{threeparttable}
\end{table}

In Figure~\ref{DECEL}, we plot the maximum Lorentz factor for the kpc-scale jet against the minimum Lorentz factor for the pc-scale jet for the sources listed in Table~\ref{decel-table}. The straight dashed line shows the case where the Lorentz factor limits of the pc-scale and kpc-scale jets are equal. The $\Gamma$ limits plotted all fall well below this line, indicating that the jets are at most mildly relativistic on the kpc scale and are experiencing significant deceleration from pc to kpc scales. These sources could only remain highly relativistic if they are very far from equipartition, which would
greatly increase the energy content in these jets.  Thus, these results suggest FR~II quasar jets decelerate at or before kpc scales, similar to what has previously been shown for FR~I radio galaxies on the basis of the jet and counter-jet intensity profiles \cite[e.g.,][]{laing_bridle_2014}.  Common mechanisms invoked to expain this jet-deceleration in FR~I radio galaxies are the entrainment of ambient cold gas by a shear boundary layer \citep[e.g.,][]{bicknell+84,deyoung86,deyoung93}, various MHD instabilities \citep[e.g.,][]{perucho+07,rossi+08,gourgouliatos+18}, and mass-loading by stars/stellar winds \citep[e.g.,][]{perucho+14,torres-alba+20,angles-castillo+21}.  

\begin{figure}
	\includegraphics[scale=0.5]{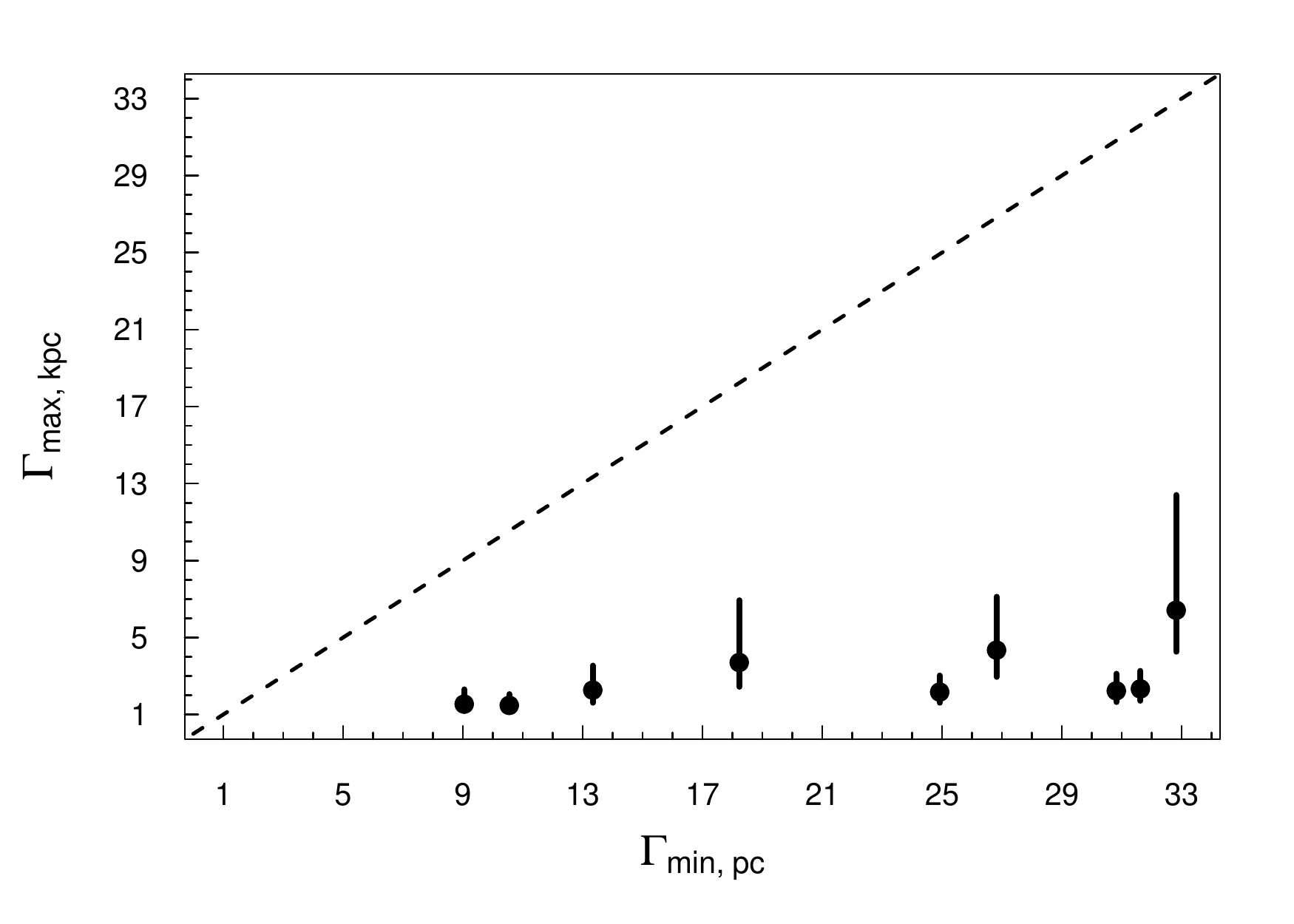}
	\caption{Plot of the maximum bulk flow Lorentz factor for the kpc-scale jet against the minimum bulk flow Lorentz factor for the pc-scale jet for the sources shown in Table~\ref{decel-table}. The black circles have error bars showing the result assuming the magnetic field can be out of equipartition by a factor of two. The dashed line shows the case where $\mathrm{\Gamma_{min,~pc}}$=$\mathrm{\Gamma_{max,~kpc}}$.  We consider points below this line evidence for deceleration.}
	\label{fig:decel}
\end{figure}

Another interpretation of this result is again motivated by a spine-sheath jet structure, as discusses in section~\ref{BEAMING PATTERNS}.  If the observed radio-to-optical synchrotron radiation originates from a slower-moving sheath surrounding a faster-moving inner spine, then our $\delta$ and $\Gamma$ upper limits for the large-scale-jet regions apply to the sheath and not the inner spine.  Thus, regardless of the X-ray emission mechanism, it is still possible the inner jet spines of the FR~IIs in Figure~\ref{fig:decel} may remain highly relativistic on kpc scales while slower boundary sheath layers dominate the observed radiative output at $\sim$GHz frequencies.

\section{Summary \& Conclusions}
\label{conclusion}

In this study, we used \textit{Fermi}/LAT observations to search for MeV-to-GeV gamma-rays predicted to result from the IC/CMB mechanism in extragalactic kpc-scale X-ray jets.  We did not find evidence for IC/CMB gamma-rays in any of the observed X-ray jets, and we subsequently ruled out the IC/CMB model in 21/45 of the X-ray jets from this study on the basis of over-predicting the observed gamma-ray flux or limits.  In the sources for which the IC/CMB model is not ruled out by our \textit{Fermi}/LAT analysis, it is possible future \textit{Fermi}/LAT observations will yield deeper limits that do rule out the IC/CMB model.  Alternatively, it is possible we will detect IC/CMB gamma-rays from some subset of these jets in the future.  Additionally, for many of these jets in which we could not rule out the IC/CMB model, we identify how future sub-mm imaging with ALMA or the SMA, or optical/IR/UV imaging with the \textit{HST} or \textit{James Webb Space Telescope (JWST)} can help determine the shape of the radio-to-optical synchrotron spectrum for better-applying our \textit{Fermi} test and measuring the limits on $\delta$.  

We have also presented additional lines of evidence against the IC/CMB model relating to the lack of correlation observed between radio and X-ray spectral indexes for the jet regions analyzed, and significant positive correlation between X-ray and radio-core dominance which has a much lower slope than required under the conventional single-zone IC/CMB model addressed in this study.  This softer correlation can be accounted for by various multi-zone emission models, where we in particular consider a two-zone synchrotron and IC/CMB model motivated by a spine-sheath jet structure.  Finally, we present evidence for pc-to-kpc-scale deceleration of a subset of our jets based upon superluminal motion measured for the pc-scale VLBI jets combined with the upper limits on the Doppler factors for the kpc-scale emitting regions analyzed in our SEDs.  The corresponding upper limits we present on the bulk-flow Lorentz factors for the kpc-scale jet regions represent a new method of constraining jet speed in kpc-scale AGN jets, and suggest kpc-scale deceleration in FR~II quasar jets.


\section*{Data Availability}

The data used in this article are all publicly available.  The \textit{Fermi}/LAT data can be found here: https://fermi.gsfc.nasa.gov/cgi-bin/ssc/LAT/LATDataQuery.cgi.  The \textit{Chandra} data can be found here: https://cda.harvard.edu/chaser/.  The \textit{HST} data can be found here: https://mast.stsci.edu/search/ui/\#/hst.  The radio interferometric data can be found here: https://data.nrao.edu/portal/\#/.



\bibliographystyle{mnras}
\bibliography{biblio} 




\appendix


\section{Spectral Models}
\label{spectral models}

\noindent \textbf{Synchrotron Fits}
\\

The empirical synchrotron fits to the radio-to-optical data are simple power laws with scaled exponential cutoffs, corresponding to a simple power-law electron energy distribution with maximum Lorentz factor.  They have the following form: 

\begin{equation}
	\mathrm{\nu f_{\nu}=N\left(\frac{\nu}{10^{10} Hz}\right)^{\gamma}exp\left(-\left(\frac{\nu}{\nu_{1}}\right)^{\beta}\right)}
\end{equation}

In this equation $\nu$ is the observed frequency of the radiation, $\gamma$ the power-law index, $\nu_{1}$ the frequency at which the exponential turnover begins, $\beta$ the steepness of the cutoff, and N is the normalization of the spectrum which has the units erg s$^{-1}$ cm$^{-2}$.  
\\

\noindent \textbf{IC/CMB shifting equations}
\\

The IC/CMB model we use has the following form for the shifting in luminosity and frequency of the lower-energy synchrotron spectrum, as first described in \cite{georganopoulos2006}:

\begin{equation}
	\mathrm{\frac{\nu_{c}}{\nu_{s}}=\frac{\nu_{CMB}}{e(B/\delta)/\left[2\pi m_{e}c(1+z)\right]}}
\end{equation}

\begin{equation}
	\mathrm{\frac{L_{c}}{L_{s}}=\frac{32U_{CMB}\left(1+z\right)^{4}}{3(B/\delta)^{2}}}
\end{equation}

B is the magnetic field strength in the emission region, e is the elementary charge, $\mathrm{U_{CMB}}$ is the CMB energy density at the current epoch, $\mathrm{\nu_{CMB}}$ the CMB peak frequency at the current epoch, z is the redshift, $\mathrm{m_{e}}$ is the electron mass, c is the speed of light, and $\delta$ is the Doppler factor.  Note the only free parameter in this shift is $B/\delta$ which becomes fixed upon fitting the X-ray component of the SED.  If one assumes an equipartition magnetic field, this shift is only parametrized by $\delta$.  Also note that these shifts should preserve the same spectral index for both the synchrotron and inverse-Compton components, assuming the extrapolation of the radio spectral index from the observed GHz frequencies to the unobserved hundreds of MHz range. 

\subsection{Robustness of Recombined Light Curve Constraints}
\label{simulation}

Our methodology for recombining \textit{Fermi}/LAT light curves in order to obtain the deepest constraints on the IC/CMB flux is based on determining the source detection significance, parametrized by the TS, in each light curve time bin and contiguously recombining the lowest significance bins until reaching flux density/upper limit minima.  In this scheme, these minima correspond to the most core quiescent time we can use in order to obtain our deepest IC/CMB constraints.  This methodology relies on the assumption that our light curves consist of a superposition of a bright, and intrisically variable, core component with a weaker and completely steady IC/CMB component.  In this setup, the source detection significance is dominated by the core component.  This allows us to use the ``core quiescent'' states when the source TS is lowest to obtain the deepest constraints on the steady IC/CMB component.  If the source consisted of only a completely steady flux component, this method of TS ordering and recombining only the lowest significance bins would lead to a biased estimate of the flux, or upper limits in the case of non-detections, since one would essentially only be using the low Poisson count deviates from random measurement error.  
	
Cursory inspection of our light curves given as online-only figures will show that they are dominated by intrinsically variabile flux likely produced by the jet cores, with complex stochastic and correlated variability patterns, flaring periods, etc.  The variability indexes published by the \textit{Fermi} team further confirm the intrinsically variable nature of the flux output from our sources \citep{4fgl}. However, in order to demonstrate that measurement bias does not significantly affect our results, we perform a simulation using the \texttt{gtobssim} tool within the \textit{Fermi} science tools software package.  This tool can simulate \textit{Fermi}/LAT events data for an assumed set of source model parameters, sky observing position, acceptance cone for your region of interest, and spacecraft pointing history.  
	
\begin{figure}
	\includegraphics[scale=0.45]{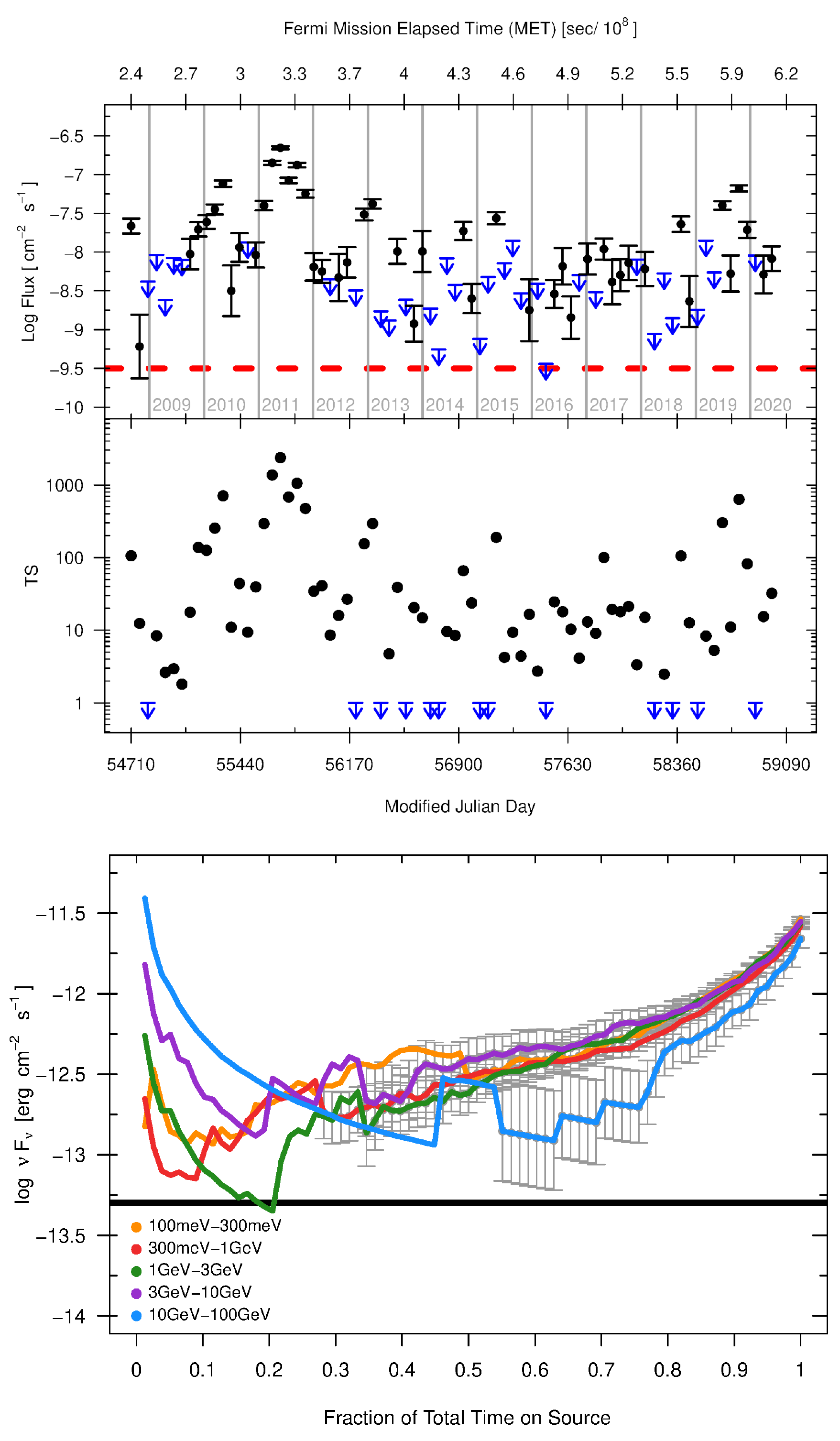}
	\caption{ Top: Simulated \textit{Fermi}/LAT light curve with corresponding TS values for each bin shown in the lower panel. The time bins correspond to three weeks in good time interval, and the flux from the source is the integrated photon flux from 100~MeV$-$100~GeV. The blue arrows in the light curve represent upper limits for bins in which the source TS was less than ten. Black data points represent fluxes measured when the TS was greater than ten, with 1$\sigma$ error bars. In the TS plot, blue arrows represent TS values less than zero, while the black points represent positive TS values. The red horizontal dashed line corresponds to the simualted steady IC/CMB flux level.  Bottom: Combined-bin flux density upper limits for the simulated light curve are shown as lines without error bars, where upper limits are measured when the source TS was less than 10. Flux densities are shown with gray 1$\sigma$ error bars on top of the lines when the source TS was greater than ten. The flux densities/upper limits are given in the five Fermi energy bands described in section~\ref{fermi_methods}.  The horizontal black lines corresponds to the simulated flux density level for the steady IC/CMB component (the red dashed line from the top panel).}
	\label{fig:simfig}
\end{figure}

For our light curve model, we simulate a steady flux component corresponding to IC/CMB, a variable component corresponding to the core, and an isotropic extragalactic background using the most recent template file, \texttt{iso\_P8R3\_SOURCE\_V3\_v1.txt}.  For our core and IC/CMB spectra we use a power-law model with photon index fixed at 2 (horizontal in $\mathrm{\nu F_{\nu}}$) for simplicity and ease of interpretation.  We base our simulation on the results for our \textit{Fermi} analysis of PKS~1150+497, using a putative source at the same sky position.  Thus, we use an isotropic background model with normalization fixed in all time bins by that found in the whole time range analysis for PKS~1150+497.  We use the actual spacecraft pointing history for our light curve time range.  We simulate core flux levels corresponding to the observed light curve fluxes for PKS~1150+497 shown in Figure~\ref{fig:lc}.  For time bins in which we report source non-detections in Figure~\ref{fig:lc}, we simulate a flux level corresponding to the flux upper limit subtracted by 0.75 dex.  These core states are fairly representative of many of our light curves, and represent a bright quasar core with intrinsic stochastic and complex variability patterns, with enough core quiescent states to provide meaningful constraints on any weak and steady flux component. Our simulated, steady, IC/CMB flux level is indicated by the horizontal red dashed line in Figure~\ref{fig:simfig}.  The steady flux level simulated for our IC/CMB component corresponds to a level $\sim$1 dex below the mean flux simulated of the core quiescent states (here we take those time bins with source non-detections in the original PKS~1150+497 light curve as the ``core quiescent states'') .  The IC/CMB simulated flux level is $\sim$ three standard deviations below the combined fluctuations of the intrinsic variability of the core quiescent states and Poisson random-error variation from a steady source with similar mean flux (with the latter being obtained from a separate simulation and both being $\sim$0.2~dex).  Thus, this still allows for time-bin-ordering by TS to be dominated by intrinsic core flux variability, even among the core quiescent states.

The results from our light curve simulation shown in Figure~\ref{fig:simfig} clearly show that, in general, the combined bin flux density/upper limit minima do not violate the steady state IC/CMB flux simulation level.  The upper limits from band~3 (3$-$10~GeV) are the only ones which just reach the simulation level, but this is consistent with statistical fluctuations from a 95\% upper limit and this does not indicate appreciable bias from our approach.

\subsection{Notes on Individual Sources}
\label{source_notes}
Next we discuss some notes on individual sources in order to highlight any features of their X-ray/radio jet morphologies, peculiararities associated with their SEDs, or other interesting source properties.  We also note the original discovery papers for each X-ray jet, any literature sources we used for constructing our SEDs, and any SED data resulting from our own analysis.  
\itemize
\item[] \textbf{J0038$-$0207 (3C~17):}  This jet is highly curved, showing a greater than 90$^{\circ}$ position angle change roughly two-thirds into its (projected) length, and commensurate with the ``bent-tail'' class of radio galaxies.  A recent study by \cite{madrid+18} shows this jet bend likely originates from the jet moving through the dense intracluster medium in which it resides, where this is a common explanation for the origin of this type of jet distortion \citep[e.g.][]{miley+72,jones+79,jones+17}.  The X-ray emission corresponds to two knots with associated radio and optical/IR emission.  One of the X-ray knots appears in the jet bend, while the other X-ray knot is a few arcseconds downstream of the core before the abrupt position angle change.  We use the radio-to-X-ray flux densities presented in the X-ray jet discovery paper \citep{massaro+09} in order to construct our SED, where we add together the flux densities for both of the observed X-ray knots.  The radio and X-ray spectral indexes are similar in value, which is consistent with the IC/CMB model.  However, the IC/CMB model is ruled out at a fairly high level of significance by our \textit{Fermi} limits, where the highest-energy \textit{Fermi} limit is closest in frequency to a region of the IC/CMB spectrum anchored by an observed portion of the synchrotron spectrum (these portions of the IC/CMB spectra correspond to where observed synchrotron flux densities would lie on the shifted IC/CMB spectra $-$ as marked by the gray circles in our SEDs).          

\item[] \textbf{J0108+0135 (PKS~B0106+013):} This jet shows semi-continuous X-ray emission along the length of the radio emission, with the bright X-ray knot B midway along its several arcsecond-long jet.  The jet morphology at both X-ray and radio wavelegnths also appears consistent with a very slight helical or S/Z shape (this S/Z shape is much more pronounced in several other of the jets presented in this study).  S/Z-shaped helical kpc-scale jets can originate from a number of processes, including: jet precession induced by a binary supermassive black hole (SMBH) \citep{begelman+80} or warped accretion disk \citep{pringle96}, various magnetohydronamic (MHD) instabilities \citep[e.g.][]{nakamura+07}, or backflowing lobe plasma \citep{leahy+84}.  The X-ray jet discovery was presented by \cite{hogan+11}, but we utilize the X-ray spectral index and flux density measurement presented by \cite{kharb+12} in our SED.  \cite{kharb+12} also note that the radio jet appears to end in a ``nose cone'' structure, believed to originate from magnetically confined jets where the plasma collects between the Mach disk and leading bow shock at the end of the jet instead of forming a reverse-shock \citep{clarke86}.  Additionally, it is suggested S/Z-shaped jet morphologies may be easier to accomodate in magnetically-dominated jets as they are easier to deflect \citep[e.g.][]{benford78}.  Our SED also makes use of archival VLA data which we imaged at L,C, and X bands, where the radio spectral index is consistent with the X-ray spectral index.  However, even after adding flux densities for all jet regions (minus the hot spot) together, the IC/CMB model is still not ruled out by the quiescent-state \textit{Fermi} flux densities (or upper limit in the case of the lowest energy band).   

\item[] \textbf{J0209+3547 (4C~35.03):}  The radio jet for this source clearly has a FR~I morphology, where the X-ray emission is associated with the brighter inner region in the first few arcseconds downstream from the core.  The X-ray jet discovery was presented in \cite{worrall+01}, where we used their measured X-ray flux density combined with radio flux densities resulting from our own multi-band VLA imaging in our SED.  This is one jet where the MSC-nature of the SED can be debated, and a single synchrotron power-law model could still be conceivably connected from radio to X-ray wavelengths  without violating any of the measurement errors. Thus, deep sub-mm-to-optical imaging is essential in order to confirm (or rule out) the MSC nature of this jet.  

\item[] \textbf{J0210$-$5101 (PKS~0208$-$512):} 
\cite{marshall+05} present the discovery of this X-ray jet,where we use the radio-to-X-ray flux densities presented in \cite{perlman+11} for our SED.  The X-ray spectral index for the jet appears harder than the radio spectral index.  Accordingly, the synchrotron and IC/CMB model fits just barely respect the associated errors.  We analyzed archival ALMA band 3 and 7 data, but could not detect the jet and thus give 95\% upper limits in our SED.  However, deeper sub-mm imaging with ALMA would be critical in order to establish the shape of the synchrotron spectrum between the measured radio flux densities and the NIR \textit{HST} jet detection.  Depending on the outcome of these observations, it is possible the IC/CMB model would remain ruled out in the event of a bright ALMA detection, or potentially lead to a SED which is broadly consistent with the IC/CMB model in the event of a non-detection or faint detection.

\item[] \textbf{J0237+2848 (4C~28.07):}  The X-ray jet discovery is reported in \cite{marshall2011}, where we use archival VLA and \textit{HST} data we reduced in our composite SED.  The radio spectral index for this source is very hard, leading to a very high level of predicted IC/CMB gamma-ray flux not observed by \textit{Fermi}.

\item[] \textbf{J0416$-$2056 (PKS~0413$-$210):} This jet exhibits a C-shaped morphology indicative of the ``wide-angle tailed'' class of radio galaxies \citep{owen+76}, where this shape may result from ram pressure experienced by the jet moving through the intergalactic/intracluster medium (similar to the bent-tail class), and potentially originating from cluster mergers \citep{sakelliou+00}.  The X-ray jet discovery was reported by \cite{marshall+05}, where we use the X-ray flux density from \cite{marshall2011} in our SED (and our own radio imaging).

\item[] \textbf{J0418+3801 (3C~111):}  This well-collimated quasar X-ray jet was first reported in \cite{hogan+11}, but we used the optical and X-ray flux densities presented in \cite{clautice2016} for our SED.  The radio flux densities in our SED came from our own analysis.  This jet is one case where the X-ray spectrum is very soft, and at odds with the steeply rising radio spectral index.  This is suggestive that the X-rays and GHz radio emission do not originate from the same electron population as is required by the IC/CMB model.  Additionally, as is apparent from our radio/X-ray imaging, the radio and X-ray knots are not cospatial in this jet as would be expected under the IC/CMB model, but rather the X-ray knots peak upstream from the radio knots (i.e., closer to the black hole).  A careful analysis of the radio/X-ray morphology of this jet is given in \cite{clautice2016}.    

\item[] \textbf{J0433+0521 (3C~120):}  This quasar was not a member of the \textit{Fermi}/LAT 3FGL 4-year point source catalog, but subsequently became detected and is a member of the \textit{Fermi}/LAT 4FGL 8-year point source catalog. The X-ray jet discovery paper is \cite{harris+99}, where we used the optical upper limits from this paper in our SED.  We also use the radio flux densities and the X-ray flux density/index from \cite{harris2004} in our SED. Knot k25 is the only X-ray knot which precludes a single synchrotron spectrum from radio-to-X-ray wavelengths.   

\item[] \textbf{J0519$-$4546 (Pictor~A):}  This is another highly collimated quasar jet, where the X-ray jet discovery was reported in \cite{wilson+01}.  The X-ray jet appears very knotty, and the radio jet is accompanied by expansive radio lobes which surround the source.  The optical jet detection was reported in \cite{gentry+15}.  We use the flux densities/upper limits and X-ray spectral index from \cite{gentry+15} for our SED.  Follow-up deep \textit{Chandra} observations were presented in \cite{hardcastle+16}, where they report the detection of jet knot flux variability and a faint X-ray counter-jet which are both at odds with the IC/CMB model (due to the fact that IC/CMB X-rays are expected to be non-variable and highly beamed).  
Our SED assumes the optical emission is the high-energy portion of a single radio-to-optical synchrotron spectrum.  It is conceiveable that IC/CMB could be rescued by instead having the optical flux density correspond to the low-energy extrapolation of the IC/CMB spectrum and appopriately softening the radio spectral index.  However, this would imply an extremely soft radio spectrum which could be easily tested with high-frequency radio observations using either ATCA or ALMA.         

\item[] \textbf{J0607$-$0834 (PKS~0605$-$085):} The X-ray jet discovery was reported in \cite{sambruna2004}, where we used their X-ray flux density and optical upper limit in our SED.  We use our own \textit{Chandra} analysis for the X-ray spectral index.  Additionally, we used our own VLA and ALMA band 3 data reduction for the SED, where the synchrotron spectrum appears to be turning over around the ALMA band 3 data point.  This turnover could be confirmed with another higher-frequency ALMA observation.

\item[] \textbf{J0728+6748 (3C~179):}  This is a knotty jet, where the X-ray detection was first reported in \cite{sambruna2004}.  We use our own archival VLA C and X-band imaging in our SED, and the X-ray/optical flux densities reported by \cite{sambruna2004}.  However, a X-ray spectral index could not be determined for the jet features analyzed, and the radio spectral index is poorly constrained by the two close-in-frequency radio flux densities.  Thus, this very northern source would make a good target for high-frequency VLA observations (i.e., K-band or above) or Submillimeter Array (SMA) observations.  Additionally, deeper \textit{Chandra} observations may provide enough photon counts to contrain the X-ray spectral index for these emission components.  

\item[] \textbf{J0741+3112 (B2~0738+31):} The X-ray jet discovery was reported in \cite{siemiginowsk+03}.  The bright X-ray knot A appears part-way along this jet and we use our own VLA images for the L/C-band flux density points in our SED.  The X-ray flux density and index used in the SED comes from \cite{sambruna2004}.  This jet is interesting because it is not necessarily a MSC jet; deep ALMA and \textit{HST/JWST} imaging could help establish the shape of the radio-to-optical synchrotron spectrum in order to confirm (or rule out) the MSC nature of this jet, and provide a more stringent test of the IC/CMB model utilizing our \textit{Fermi} limits.  

\item[] \textbf{J0830+2410 (B2~0827+24):}  This is another bent-jet source, where the radio emission appears to be stronger downstream from the jet bend, and the X-ray emission peaks upstream from the bend.  The X-ray jet discovery was reported in \cite{jorstad+04}. 

\item[] \textbf{J0840+1312 (4C~13.38):}  The X-ray jet discovery was reported in \cite{sambruna2004}, where use knot A for the SED since this knot has an associated optical detection.  The X-ray spectral index appears significantly harder than the radio spectral index for this source, where our radio spectrum is very well-constrained by our own multi-band VLA and ALMA imaging (we use the C-band flux density from \cite{sambruna2004}).  The optical/X-ray flux densities and X-ray spectral index are taken from \cite{sambruna2004}.  

\item[] \textbf{J0922$-$3959 (PKS~0920$-$397):} The X-ray jet discovery is reported in \cite{marshall+05}, where we use the X-ray flux density from \cite{schwartz+06} and our own \textit{Chandra} analysis for the X-ray spectral index measurement.  We also use our own ATCA and ALMA data reduction in order to fill out the radio synchrotron spectrum.  This is one case where we used our own archival \textit{HST} analysis and were able to detect the entire optical jet.  The SED is constructed using all of the emission components in the jet without including the hot spot.   One obvious feature of this jet's morphology is its quasi-periodic ``cannonball'' knots, where one interpretation for this morphology is that it results from the periodic modulation of the accretion rate due to the dynamics of a binary SMBH \citep[i.e.][]{godfrey2012}.        

\item[] \textbf{J0947+0725 (3C~227):}  This X-ray jet was first reported in \cite{hardcastle+07}, where only one X-ray feature was reported as potentially corresponding to a jet knot versus just a hot spot: P4.  Thus, our SED analyzes this feature, but our \textit{Fermi} limits are not constraining enough to rule out the IC/CMB model.  We use the X-band flux density from \cite{hardcastle+07} combined with our own measured L/C-band flux densities in our SED.  

\item[] \textbf{J1001+5553 (QSO~0957+561):}  This quasar system is gravitationally lensed, where the lensing results in a double image of the quasar.  The lensing and X-ray jet discovery are presented in \cite{chartas+02}, where we also use the X-ray spectral index and flux density from this paper in our SED.  We use our own data reduction for the radio flux densities.  This is a case where the jet X-ray spectral index is significantly softer than the radio spectral index.  However, in this case the X-ray index is measured for the entire jet while our SED is constructed just for knot B.  Therefore, it is possible the X-spectrum of just knot B is more in line with the IC/CMB model.  Higher-frequnecy radio observations are critical for this jet in order to appropriately test the IC/CMB model.

\item[] \textbf{J1007+1248 (4C~13.41):} This X-ray jet was first reported in \cite{miller+06}, where they also discuss the observed broad absoption lines in the optical/UV spectra originating from a fast gaseous outflow in this system (and indicative of jet-driven feedback with its host galaxy).  The radio jet hosted by this quasar is enormous in size, falling just shy of the 0.7~Mpc cutoff for the ``giant radio galaxy'' class \citep[e.g.][]{lara+01,schoenmakers+01}, coming in at $\sim$0.57~Mpc in linear projected size.  The twin radio jets hosted by 4C~+13.41 have a hybrid morphology \citep{gopal+00}, where the southeastern component has a FR~I-type plumey jet, and the northwestern jet exhibits conventional edge-brightened FR~II-type features.  Interestingly, the anomalousl X-ray emission in this MSC jet is from the FR~I side and not the FR~II side, where anomalous X-ray emission is usually observed in FR~II and not FR~I jets.  The origin of hybrid morphology radio galaxies is widely debated, with possible explanations including an assymetric environment surrounding the AGN (i.e., the FR~I jet was FR~II-type jet disrupted by propagating into a relatively dense IGM) \citep{meliani+08}, projection/Doppler-boosting effects leading to the illusion of hybrid morphology \citep{degasperin17}, or an illusion caused by a combination of light-travel time effects and retriggered AGN activity \citep[e.g.,][]{marecki12}.

\begin{figure}
	\includegraphics[scale=0.55]{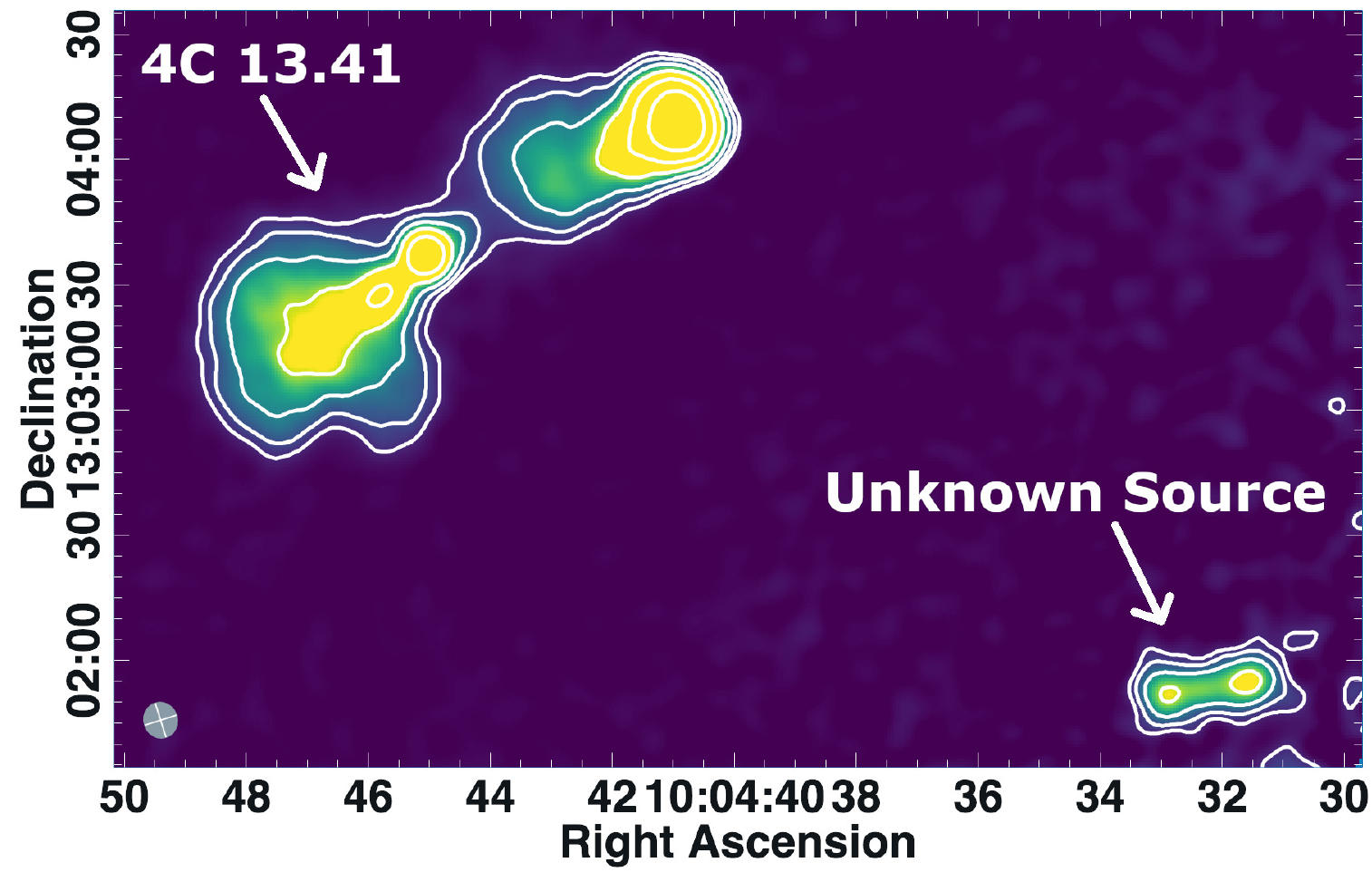}
	\caption{Here we present the VLA X-band (8.46~GHz) image for 4C~13.41 (with RA/DEC given in the FK4/B1950 coordinate system), showing the serendipitous radio source discovered at the edge of our field in the bottom right corner of the image.  The serendipitously discovered source appears consistent with a conventional FR~II radio galaxy viewed with its jets oriented perpendicular to the line of sight and inflating diffuse lobes on either side of the black hole.  The radio contours start at a base level of 3$\sigma$, and increase by factors of two thereafter.  The image intensity color map ranges from 0 to 5.6~$\mathrm{mJy\ beam^{-1}}$.  The synthesized beam is shown in the bottom left as a solid gray ellipse with a cross.}
	\label{fig:4c13_41}
\end{figure}

We use our own C/X-band VLA imaging for the SED (in addition to ALMA upper limits based upon non-detections), combined with the X-ray index and flux density presented in the discovery paper.  For this jet, higher-frequency and deep VLA or ALMA imaging is essential to better constrain the radio-to-optical synchrotron spectrum.  In our deep X-band radio photometry we also serendipitously detected an unknown radio source which we could not find a match for in existing AGN and galaxy catalogs.  We present the X-band image showing this radio source in Figure~\ref{fig:4c13_41}.  This very nearby radio galaxy is suggestive that the two systems may reside in a dense galaxy cluster, where radio-loud AGN are generally found to have a preference for residing in cluster envrionments \citep[e.g.,][]{shen+09}.  This picture supports the scenario that frequent galaxy mergers might be necessary to help ignite radio-jet activity in AGN \citep[e.g,][]{chiaberge15}.  

\item[] \textbf{J1033$-$3601 (PKS~1030$-$357):}  The X-ray jet discovery for this quasar was first presented in \cite{marshall+05}, where we use the measurements presented in \cite{schwartz+06} for the X-ray flux density and optical upper limit.  The radio flux densities and X-ray spectral index come from our own analysis.  This is one case where deep ALMA imaging would help establish the shape of the radio-to-optical synchrotron spectrum for our \textit{Fermi} test.  

\item[] \textbf{J1048$-$1909 (PKS~1045$-$188):} This jet is an example of a slightly bent jet, where the X-ray detection was first reported by \cite{hogan+11}.  We use the X-ray flux density/spectral index presented in \cite{stanley15} for our SED, where the radio-to-optical flux densities result from our own archival data analysis.  

\item[] \textbf{J1048$-$4114 (PKS~1046$-$409):}  This X-ray jet was first reported in \cite{marshall+05}.  This jet shows a strong bend, with a majority of the X-ray flux emanating from the jet upstream to this bend and downstream of the core.  There is also a very bright feature northwest of the core which is itself associated with a lobe-type structure.  This feature has a two-point 8.64~GHz$-$19~GHz spectral index of $\alpha~\sim~0.6$, which is more consistent with a jet knot than a core.  For this reason we don't believe it is a dual AGN, or an unrelated background/foreground AGN.  We use the X-ray flux density given in \cite{marshall+05} for our SED, but our own imaging for the radio flux densities.  However, this SED is poorly constrained; deep \textit{Chandra} observations would be beneficial to nail down the X-ray spectral index along with radio follow-up to better constrain the radio spectral index and shape of radio-to-optical synchrotron spectrum.  In this vain, deep optical/IR imaging with the \textit{HST/JWST} would help anchor the radio-to-optical synchrotron spectrum.      

\item[] \textbf{J1058+1951 (4C~20.24):} The discovery of this X-ray jet was first presented in \cite{schwartz+06_1055}, where we use the X-ray spectral index and flux density in this paper for our SED.  We use our own L/C/X-band VLA and ALMA band 3/6 imaging to fill out the radio synchrotron spectrum, which is very well constrained.  The X-ray spectrum is very soft, and highly discrepent with the radio spectral index.  Interestingly, the X-ray emission is very continuous and not knotty like some of the other X-ray jets presented in this study.  A very slight jet bend is apparent in this source.    

\item[] \textbf{J1130$-$1449 (PKS~1127$-$145):}  The X-ray jet discovery for this source was first reported in \cite{siemiginowska+02}.  We use the X-ray flux density/spectral index presented in \cite{siemiginowska+07} in our SED.  Our radio flux densities come from our own analysis.  This is a case where a single synchrotron spectrum can not be definiteively ruled out from radio-to-X-ray frequencies.  Thus, deep ALMA and/or \textit{HST/JWST} imaging could help to confirm (or refute) the MSC nature of this jet and also better-constrain the shape of the radio-to-optical synchrotron spectrum.  

\item[] \textbf{J1153+4931 (PKS~1150+497):} This X-ray jet was first reported in \cite{sambruna02}, where we use the radio-to-X-ray flux densities and X-ray spectral index presented in \cite{sambruna06} for our SED.  This jet is another case of a helical jet with quasi-periodic cannonball knot structure $-$ both potential indicators of a binary SMBH system.  We show the SED for the brightest X-ray knot (knot B), where we rule out the IC/CMB model at a high level of significance and the radio-to-optical synchrotron spectrum is very-well constrined.     

\item[] \textbf{J1205$-$2634 (PKS 1202$-$262):}  This is another case of a curved jet, where the X-ray jet discovery is reported by \cite{marshall+05} and we use the X-ray flux density and \textit{HST} upper limits reported in \cite{perlman+11} for our SED.  We use our own analysis for the X-ray spectral index measurement.  The X-ray emission in this case seems to be continuous along the length of the jet.  We use our own radio imaging for the radio flux densities, where there is an apparent synchrotron spectral turnover by the ALMA band 3 and 6 data points. Additionally, there is a definite tension between the radio spectral index and X-ray spectral index, with the X-ray spectral index being softer than the radio.   

\item[] \textbf{J1224+2122 (4C~21.35):} This X-ray jet was first reported by \cite{jorstad2006}, where we use their X-ray flux density in our SED.  This jet is a clear example of a S/Z-shaped jet, with quasi-periodic knot structure suggestive of a binary SMBH.  The first two jet knots have associated X-ray emission (where we combine the flux densities of these knots for our SED), while the rest of the jet appears undetected in X-rays.  We have an extensive amount of radio imaging we used to fill out the radio-to-optical synchrotron spectrum.  However, this is still a source where even higher-frequency (i.e., greater than ALMA band 6) ALMA imaging would be helpful to determine if IC/CMB is ruled out or if it is consistent with the broadband SED.  We show an ALMA band 3 in-band spectral index which appears consistent with the broadband radio spectral index, and predicts a rising flux density for future high-frequency ALMA observations.  Another way to anchor the radio-to-optical synchrotron spectrum would be to pursue deep \textit{HST/JWST} imaging in the hopes of an optical jet detection.        

\item[] \textbf{J1319+5148 (4C~52.27):}  The X-ray jet discovery was reported in \cite{jorstad2006}, where we use their measured X-ray flux density in our SED.  The radio jet has a knotty morphology and appears  to end in a nose-cone structure.\ \ 
Interestingly, the X-ray emission in this jet originates from the knot (knot E) upstrem of the nose cone, consistent with the idea that high-energy particle acceleration is not occuring in the nose cone due to a lack of shocks in this region.  We used our own C/Ku-band VLA imaging for our SED, but higher-frequency radio observations and \textit{HST/JWST} observations are essential to better-test the IC/CMB model in this source.  Furthermore, this jet is another case where the MSC nature has not been definitively established; deep SMA and/or \textit{HST/JWST} observations could help in this regard.     

\item[] \textbf{J1325+6515 (4C~65.15):}  The X-ray jet discovery for this quasar was reported in \cite{miller+09}, where we use the X-ray flux density/index and \textit{HST} upper limit presented in this paper for our SED. Our VLA L/X/Ku-band imaging appears to just detect the syncrhotron spectrum as it starts to turn over.  Thus, low frequency (i.e., lower than L-band) VLA observations, high-frequency observations with the SMA, and deep \textit{HST/JWST} imaging would be beneficial to better characterize the radio-to-optical synchrotron spectrum.  This radio jet is classified as a hybrid morphology jet, with one side displaying a FR~I morphology and the other a FR~II morphology.  Interestingly, the anomalous X-ray emission for this jet appears spatially cocincident with the (drastic) jet bend in this source, and like 4C~13.41 occurrs on the FR~I jet side.  

\item[] \textbf{J1421$-$0643 (PKS~1418$-$064):}  This X-ray jet is hosted by a very high redshift quasar (z = 3.7), and its discovery was reported in \cite{mckeough+16}.  We use the X-ray flux density/spectral index and \textit{HST} upper limits presented in \cite{worrall+20} for our SED.  We use our own VLA L/C-band images for the radio flux densities presented (along with upper limits from two archival ALMA non-detections).  Deeper ALMA and \textit{HST/JWST} observations are essential for establishing the MSC nature of this jet and the shape of the radio-to-optical synchrotron spectrum for our \textit{Fermi} test.  This X-ray jet is an especially important test for the IC/CMB model owing to its extremely high redshift, and the fact that the IC/CMB mechanism is supposed to be much more efficient with increasing redshift due to $\mathrm{(1+z)^{4}}$ dependance of the CMB energy density on redshift \citep[e.g.,][]{schwartz02,marshall+18}. 

\item[] \textbf{J1510+5702 (QSO~B1508+572):}  This is the highest-redshift quasar in our sample (z = 4.3), and the discovery of its X-ray jet was reported by \cite{yuan+03} and \cite{siemiginowska_1508+03}.  We use the L-band radio flux density reported by \cite{cheung+04} combined with a Ku-band flux density resulting from our own VLA analysis (including a core subtraction) in order to constrain the radio synchrotron spectrum.  We also use the \textit{HST} upper limit and X-ray flux density/spectral index presented in \cite{yuan+03} in our SED.  Significantly, the IC/CMB model is ruled out in this very high redshift X-ray jet where one would expect the IC/CMB mechanism to be much more efficient.  This quasar is also another source which was not detected in the \textit{Fermi}/LAT 3FGL 4-year point source catalog, but became a detection in the 4FGL 8-year point source catalog.  

\item[] \textbf{J1632+8232 (NGC~6251):} This X-ray jet discovery was reported by \cite{sambruna+04_ngc6251} and \cite{evans+05}, and the jet is hosted by a very nearby massive elliptical galaxy.  The discovery papers also note a detection of a diffuse X-ray-emitting gaseous halo (of unknown origin) and cavities excavated in said halo by the radio-luminous jet, indicative of jet-mode feedback operating in this system (on the scale of the cluster).  The X-ray emission associated with the outer jet region in this source is anomalous, and is the region for which we constructed a SED.  We use the X-ray flux density/spectral index from \cite{evans+05} in our SED, along with our own X/L-band VLA imaging.  Critically, deep SMA and/or \textit{HST/JWST} imaging would help to establish the shape of the radio-to-optical synchrotron spectrum in this jet for a more suitable test of the IC/CMB model with \textit{Fermi}.  

\item[] \textbf{J1642+3948 (3C~345):} This X-ray jet discovery was reported by \cite{sambruna2004}, where we use the X-ray flux density/spectral index and \textit{HST} upper limit from \cite{kharb+12} in our SED.  We use our own analysis for the radio flux densities.  Deep ALMA observations could help determine the shape of the radio-to-optical synchrotron spectrum for best-applying our \textit{Fermi} test.   

\item[] \textbf{J1642+6856 (4C~69.21):}  This X-ray jet was first published in \cite{sambruna2004}, where we used their measured X-ray/optical flux densities in our SED.  We used our own \textit{Chandra} analysis to determine the X-ray spectral index for the jet, and used our own VLA C/X/Ku-band imaging for the radio flux densities.  We combine the  flux densities from all of the jet knots together for our  SED, but the brightest X-ray knot is roughly two-thirds down the length of the slightly curved radio jet (knot C).        

\item[] \textbf{J1720$-$0058 (3C~353):}   The discovery of this X-ray jet was reported in \cite{kataoka+08}, where we use their X-ray flux density/X-ray spectral index in our SED.  We also use the three lowest-frequency radio flux densities from \cite{kataoka+08} combined with our own analysis of X and K-band VLA data for the two highest-frequency radio points.  Our SED focuses on the brightest X-ray knot in the jet, and this SED is a clear example where the radio spectral index is significantly harder than the X-ray spectral index.  As noted in the discovery paper, this source also has a X-ray counter jet, inconsistent with a highly aligned jet as is required under the IC/CMB model (where we would not expect to detect a counter jet due to the X-rays being relativstically beamed out of our line of sight).

\item[] \textbf{J1746+6226 (4C~62.29):}  This is another case of a high-redshift quasar (z = 3.9).  The radio jet morphology resembles the knotty quasi-periodic cannonball jets, with a nose-cone-like jet termination.  The X-ray emission appears continuous along the length of the jet, ending just before the jet termination feature.  The X-ray jet discovery was first reported in \cite{cheung+06}, where we use their X-ray spectral index/flux density and \textit{HST} upper limit in our SED.  We use our own VLA imaging for the SED, including an in-band spectral index bowtie on top of the Ku band flux density.  It appears our VLA flux densities are probing the synchrtron spectrum just at its turnover for this source.  It seems unlikely future \textit{Fermi} observations will be able to rule out the IC/CMB model in this jet.      

\item[] \textbf{J1829+4844 (3C~380):}  This jet is only $\sim$10~kpc in linear (projected) extent, where the X-ray and radio emission seem roughly coincident along the length of this small jet.  The X-ray jet discovery was reported in \cite{marshall+05}, where we use the X-ray flux density from this paper combined with the radio/optical flux densities presented in \cite{odea+99}.  The IC/CMB model is only marginally ruled out in this jet from our highest energy \textit{Fermi} limit, but future \textit{Fermi} observations may be able to increase the significance of this gamma-ray non-detection.  

\item[] \textbf{J1849+6705 (8C~1849+670):}  The X-ray jet discovery for this source is reported by \cite{hogan+11}, but we used the X-ray flux density/spectral index and \textit{HST} upper limits from \cite{stanley15} in our SED.  We used our own VLA L/X-band data reduction for the radio flux densities.  This appears to be another S/Z-shaped jet, where the X-ray emission is associated  with the inner jet before it starts to bend.  Sub-mm imaging with the SMA could help determine the synchrotron spectrum at a frequency which would better test the IC/CMB model with our \textit{Fermi} analysis.      

\item[] \textbf{J1927+7358 (4C~73.18):}  The X-ray jet discovery was reported in \cite{sambruna2004}, where we use their \textit{HST} flux density in our SED.  We use our own radio/X-ray analyses for our other flux densities and X-ray spectral index measurement.  The X-ray emission is associated with the first inner knot (knot A).  This is also another example of a bent-tail radio galaxy.  The X-ray emission is very soft, but not inconsistent with the measured radio flux densities.  The X-ray emission is only marginally anomalous, where the optical emission appears likely to belong to the same spectral component as the X-rays. A single synchrotron model from radio to X-rays might be able to fit the data, all be it at a slight tension with the measurement uncertainties.   Deep SMA imaging would help give us a better understanding of the radio-to-optical synchrotron spectrum in this jet.        

\item[] \textbf{J2005+7752 (S5~2007+777):}  This is a very knotty and hard, resolved X-ray jet hosted by a BL~Lac object $-$ first published in \cite{sambruna+08}.  We use all of the flux densities, \textit{HST} flux density upper limits, and the X-ray spectral index presented in \cite{sambruna+08} for our SED.  As is apparent from the SED, higher-frequency radio observations would be helpful in establishing the shape of the radio-to-optical synchrotron spectrum (where the highest-frequency point in our SED is at 4.86~GHz).  This jet is another case of a hybrid morphology jet, with the X-ray emission corresponding to the FR~I side.     

\item[] \textbf{J2105$-$4848 (PKS~2101$-$490):}  This X-ray jet was first reported in \cite{marshall+05}, where we use the multi-wavelength flux densities and X-ray spectral index presented in \cite{godfrey2012_pks2101} for our SED.  We also analyzed ALMA band 3 and 6 data for our SED, where these flux densities suggest we may be seeing a radio spectral turnover around ALMA band 3.  The radio morphology for this jet is very knotty, and displays a slight curve.  

\item[] \textbf{J2158$-$1501 (PKS~2155$-$152):} The X-ray jet discovery was reported in \cite{hogan+11}, where we used the X-ray flux density from this paper in our SED.  We used our own \textit{Chandra} analysis for the X-ray spectral index, and our own radio/optical analysis for the radio flux densities (including the upper limits from the ALMA non-detections) and \textit{HST} flux density upper limit.  Due to the deep ALMA band 6 upper limit, it's unlikely \textit{Fermi} will ever be able to rule out the IC/CMB model in this jet.      

\item[] \textbf{J2203+3145 (4C~31.63):}  This quasar is interesting since it was in the 3FGL catalog, but then was not detected in the 4FGL catalog.  Additionally, we find no evidence for a kpc-scale, resolved X-ray jet in our \textit{Chandra} analysis of this source.  However, the X-ray jet was first reported in \cite{hogan+11}.  The IC/CMB model is not ruled out by our \textit{Fermi} limits.  We used the L-band flux density  and X-ray flux density from \cite{hogan+11} in our SED. We used our own X-band analysis for the higher-frequency radio flux density.     

\item[] \textbf{J2218$-$0335 (PKS~2216-038):}  The X-ray jet discovery was published in \cite{hogan+11}.  The radio flux densities and X-ray spectral index measurement used in our SED come from our own analysis.  Additionally, we used the X-ray flux density and \textit{HST} upper limits from \cite{stanley15} in our SED.  The X-ray and radio jets are very knotty, and also show a helical pattern.  We show an ALMA non-detection flux density upper limit from our own analysis which is not constraining at all.  Deeper ALMA imaging would help constrain the radio-to-optical synchrotron spectrum in this case.      

\item[] \textbf{J2253+1608 (3C~454.3):}  This source is another blazar feautring resolved, kpc-scale X-ray emission \citep[X-ray jet discovery paper:][]{marshall+05}.  We use the optical flux densities from \cite{tavecchio+07} and X-ray flux density/index from \cite{marshall+05} in our SED. We use our own radio imaging for the SED, where we appear to be detecting the radio spectrum just past the spectral turnover.  For this reason, it is not surprising that our ALMA analysis was unable to detect the jet, and we therefore show flux density upper limits in the sub-mm/mm bands.  This is another case  of a very knotty, cannonball jet.    

\item[] \textbf{J2338+2701 (3C~465):}  This X-ray jet discovery was reported in \cite{hardcastle+05}, and the radio jets resemble the morphology of the wide-angle tail radio galaxy class.  The X-ray emission occurs in the inner portion of the radio jet upstream of one of the flaring tails.  We use the X-ray flux density reported in \cite{hardcastle+05} in our SED and our own VLA/ALMA imaging for the radio flux densities.  We constrcuted a SED for Knot D (the brightest X-ray knot in the jet). 

\subsection{Jet X-ray/Radio Images \& Broadband SEDs}
\label{images_SEDs}

Here we show the radio/X-ray morphologies for our sample, with false-color X-ray images overlaid with radio contours.  The radio data used to make these images is designated in Table~\ref{radio-table}.  The radio contours  start at a base level of 5$\sigma$ and increase by factors of two thereafter (RMS values are given in Table~\ref{radio-table}).  Our \textit{Chandra} images are also smoothed with a Gaussian of varying radii in order to better illustrate the jet morphology.  All jet regions analyzed for our SEDs are labeled in the images, in addition to other regions of interest (the regions analyzed for our SEDs are noted in the SED figures and in Table~\ref{fermi-table}).

Our SEDs shown are for the jet regions analyzed in order to test the IC/CMB model.  In most cases these SEDs correspond to the brightest (anomalous) X-ray knots, but in some cases we used the entire X-ray jet or combined data from several knots.  It's important to note that data from several knots can be added together for this analysis since the \textit{Fermi} limits (or minimum flux densities) apply to the entire large-scale jet.

\begin{figure*}
	\includegraphics[scale=0.2]{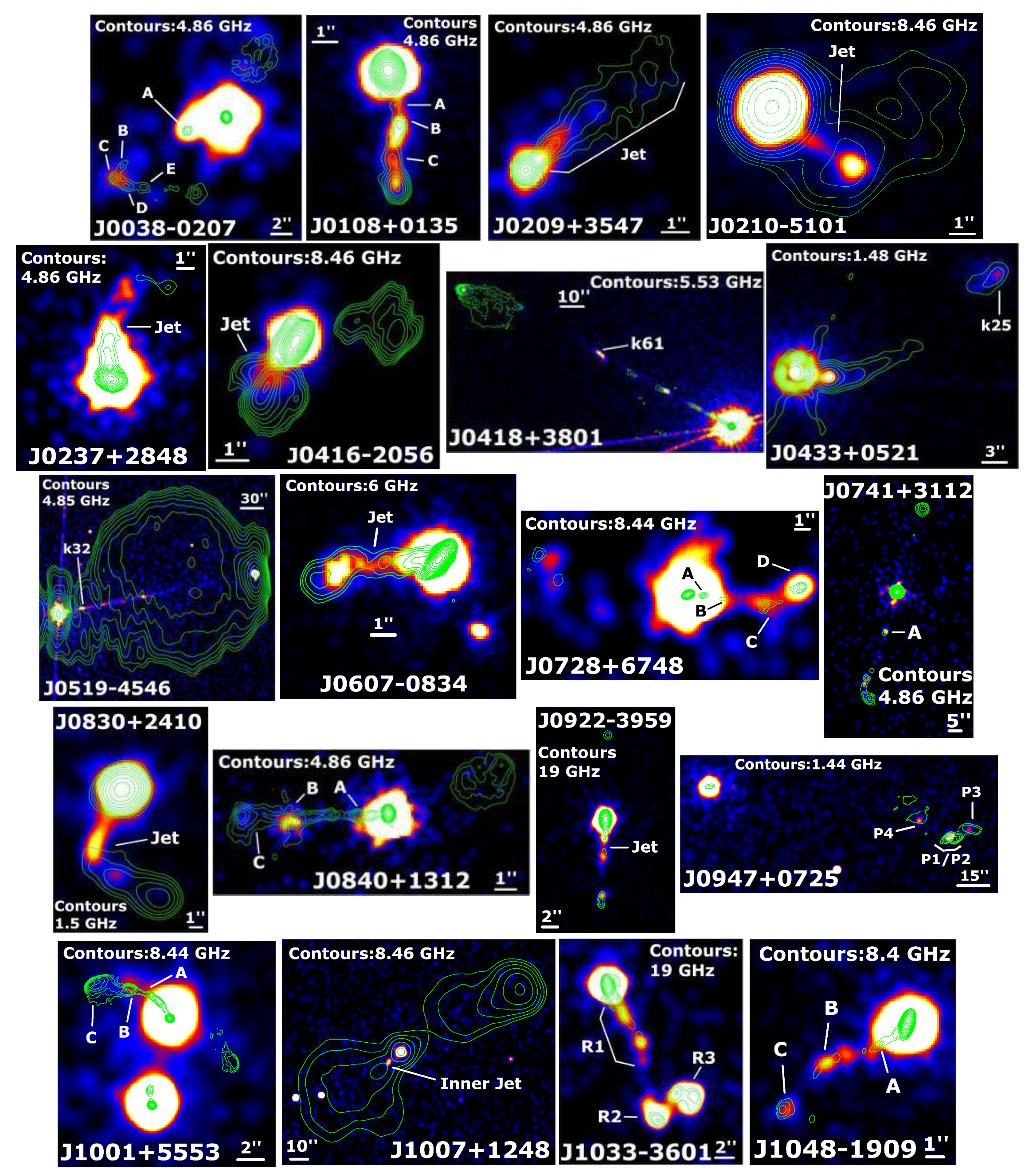}
	\caption{\textit{Chandra} X-ray images of our jet sample are shown as false-color images on a linear scale with the ``b'' color map of SAOImageDS9.  
	The color map used ranges from black to white, with warmer colors correpsonding to higher pixel values.  The X-ray images are smoothed with Gaussian kernals of various radii. 
	Overlaid on these X-ray images are radio contours, where the archival data used to make these images is described in Table~\ref{radio-table}.  The radio contours are at a base level of 5$\sigma$, spaced by factors of two thereafter (with RMS and synthesized beam properties also given in Table~\ref{radio-table}).  The jet regions analyzed for the SEDs are delineated in the images, though in some cases we also label other jet components (see the SEDs or Table~\ref{fermi-table} for clarity on which region was analyzed for the IC/CMB \textit{Fermi} test).}
	\label{fig:img1}
\end{figure*}

\begin{figure*}
	\includegraphics[scale=0.22]{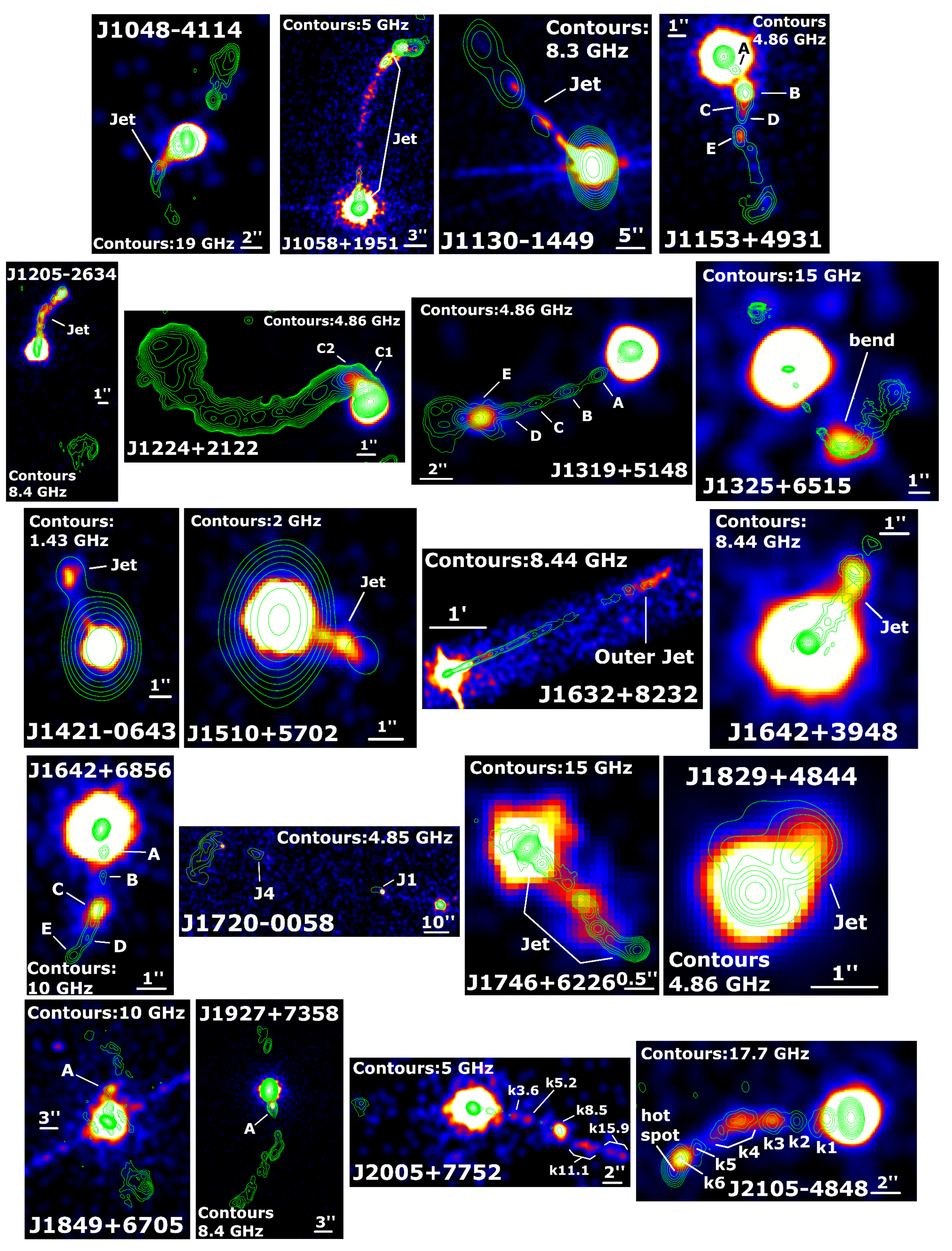}
	\caption{This is a continuation of Figure~\ref{fig:img1}}
	\label{fig:img2}
\end{figure*}

\begin{figure*}
	\includegraphics[scale=0.2]{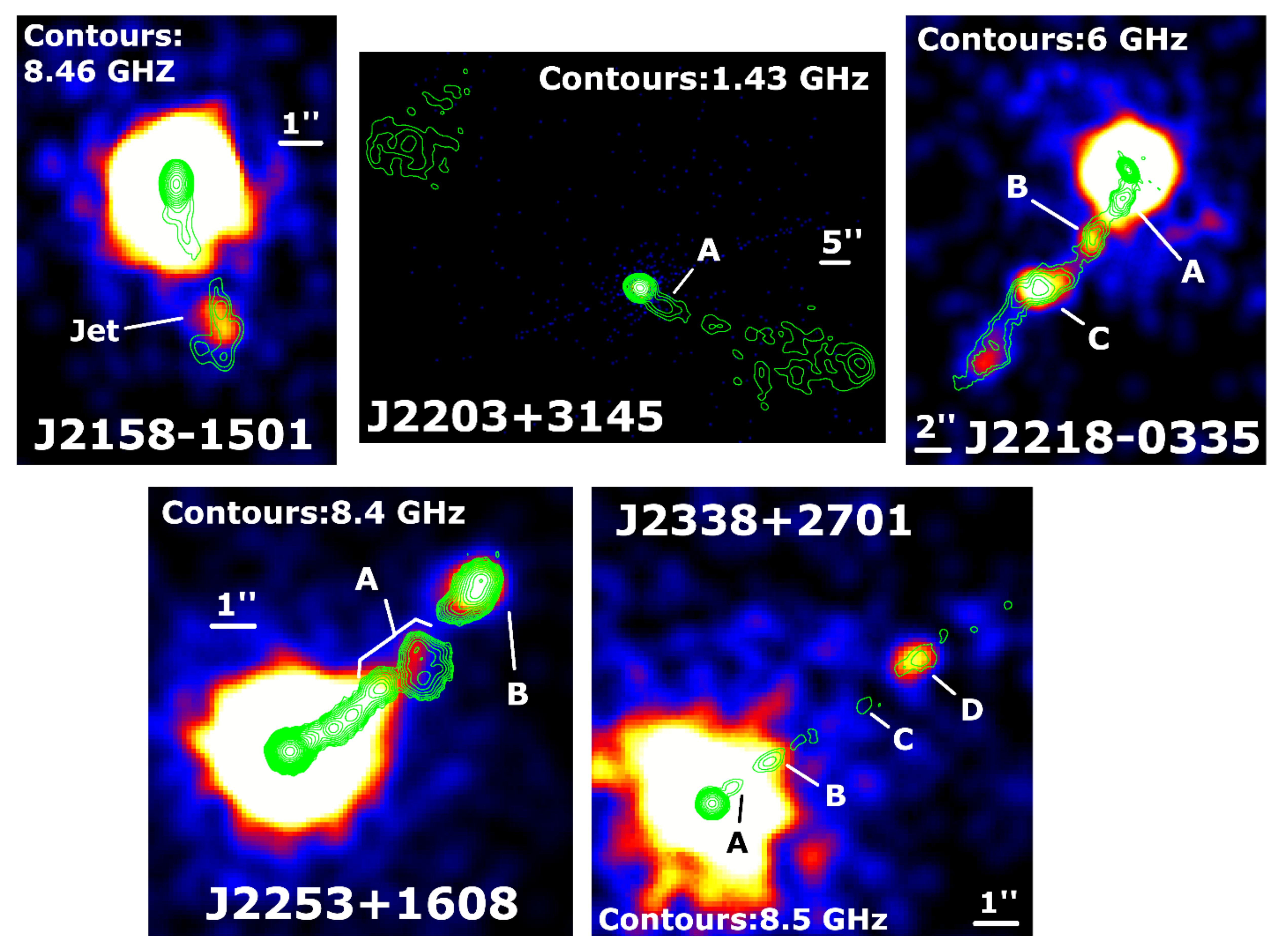}
	\caption{This is a continuation of Figures~\ref{fig:img1} and \ref{fig:img2}.}
	\label{fig:img3}
\end{figure*}

The flux densities used in these SEDs were often taken from the literature, except for cases in which we used our own analysis (as described on a source-by-source basis in section~\ref*{source_notes}). Bowties are shown to illustrate the spectral indexes with associated uncertainties, and are also taken from the literature when possible.  When X-ray spectral indexes were not available in the literature, we used our own \textit{Chandra} analysis when possible to measure the X-ray spectral indexes.  In a few cases we also show in-band radio spectral indexes as bowties in order to better constrain any potential spectral turnovers of the synchrotron spectrum.  We show the phenomenological synchrotron fits as thin black lines, and the IC/CMB model fits as thick black lines (details concerning these spectral models can be found in section~\ref{spectral models}).  We used radio spectral indexes which best-matched the measured X-ray spectral indexes, as the radio and X-ray spectral index should match under the IC/CMB model.  We also used synchrotron spectral turnovers which either matched the optical flux for optical detections or were the most conservative in the case of optical non-detections.  Gray points mark where the measured radio-to-optical synchrotron flux densities would appear on the shifted IC/CMB spectra.  The minimum flux densities (or flux density upper limits) for the large-scale jet gamma-rays observed by \textit{Fermi}/LAT are given in red, following the methodology presented in Section~\ref{fermi_methods}.  The naming convention for our jet components analyzed may differ from other works, so it is important to reference the radio/X-ray images (Figures~\ref{fig:img1}-\ref{fig:img3}) in interpreting these SEDs.  The dotted lines illustrate our lack of knowledge for the high-energy radio-to-optical synchrotron spectrum in cases where this information is critical in the determination of whether the IC/CMB model is ruled out.  The dashed lines correspond to the IC/CMB model fits corresponding to the dotted synchrotron spectra.  It's important to note that for the purposes of determinining the status on whether or not IC/CMB is ruled out and the value of the $\delta$ limits, we use the conservative dashed-line IC/CMB model predictions.

\begin{figure*}
	\includegraphics[scale=0.33]{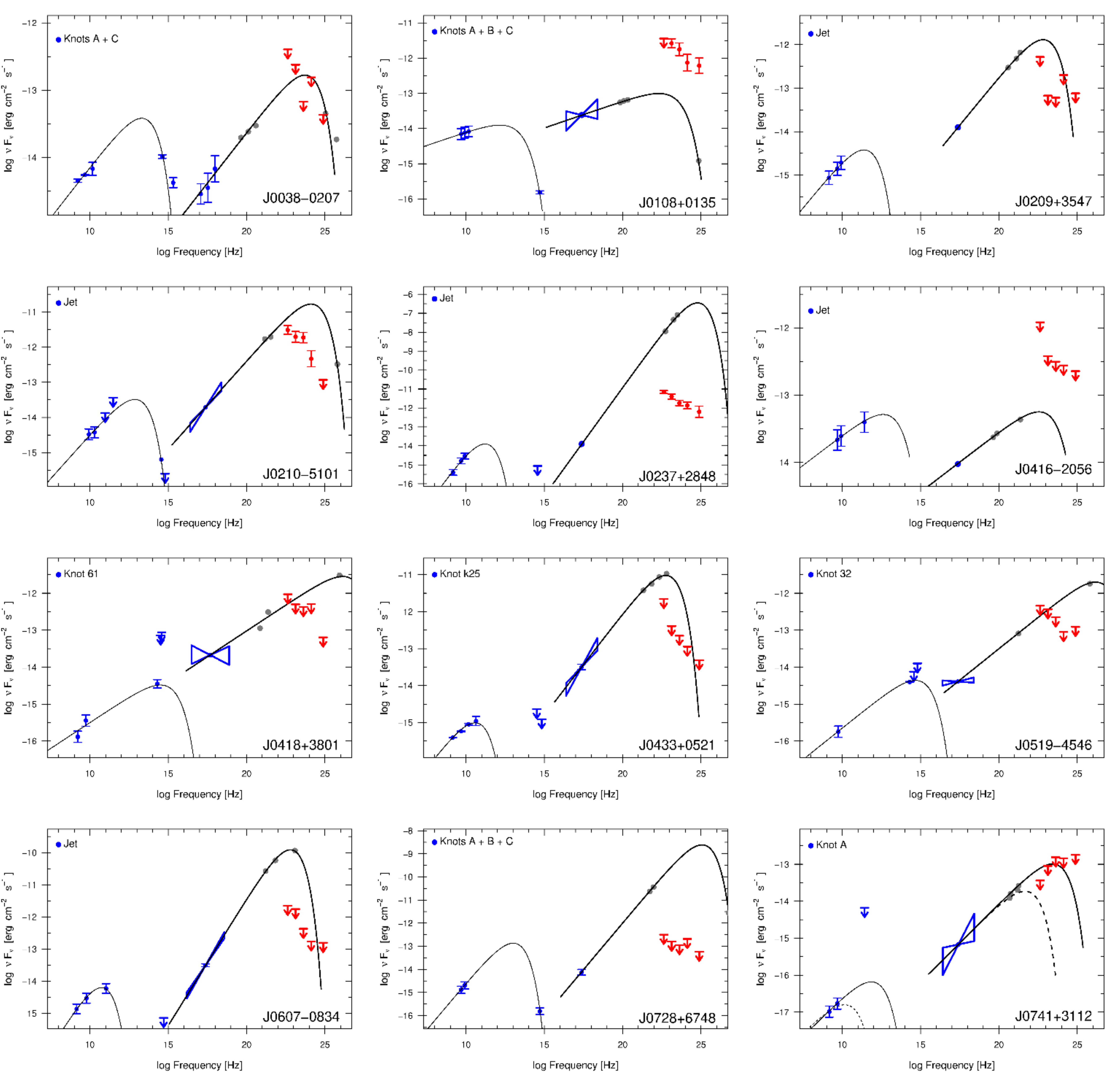}
	\caption{SEDs are shown for each jet from our sample, with blue data points corresponding to measured flux densities with associated 1$\sigma$ error bars.  These flux densities were taken from the literature, except in a few cases in which we use our own analysis (this is described on a source-by-source basis in section~\ref{source_notes}). Bowties are shown to illustrate spectral indexes with associated uncertainties, which are also taken from the literature except for a few sources in which we used our own analysis (all radio spectral index bowties result from our own data analysis).  We show the phenomenological synchrotron fits as thin black lines, and the IC/CMB model fits as thick black lines.  Gray points mark where the measured radio-to-optical synchrotron flux densities would appear on the shifted IC/CMB spectra and can be considered anchor points for the IC/CMB spectra.  The minimum flux densities (or flux density upper limits) for the large-scale jet gamma-rays observed by \textit{Fermi} are given in red, following the methodology presented in Section~\ref{fermi_methods}.  The dashed lines are IC/CMB model fits corresponding to the dotted synchrotron spectra.}
	\label{fig:sed1}
\end{figure*}

\begin{figure*}
	\includegraphics[scale=0.33]{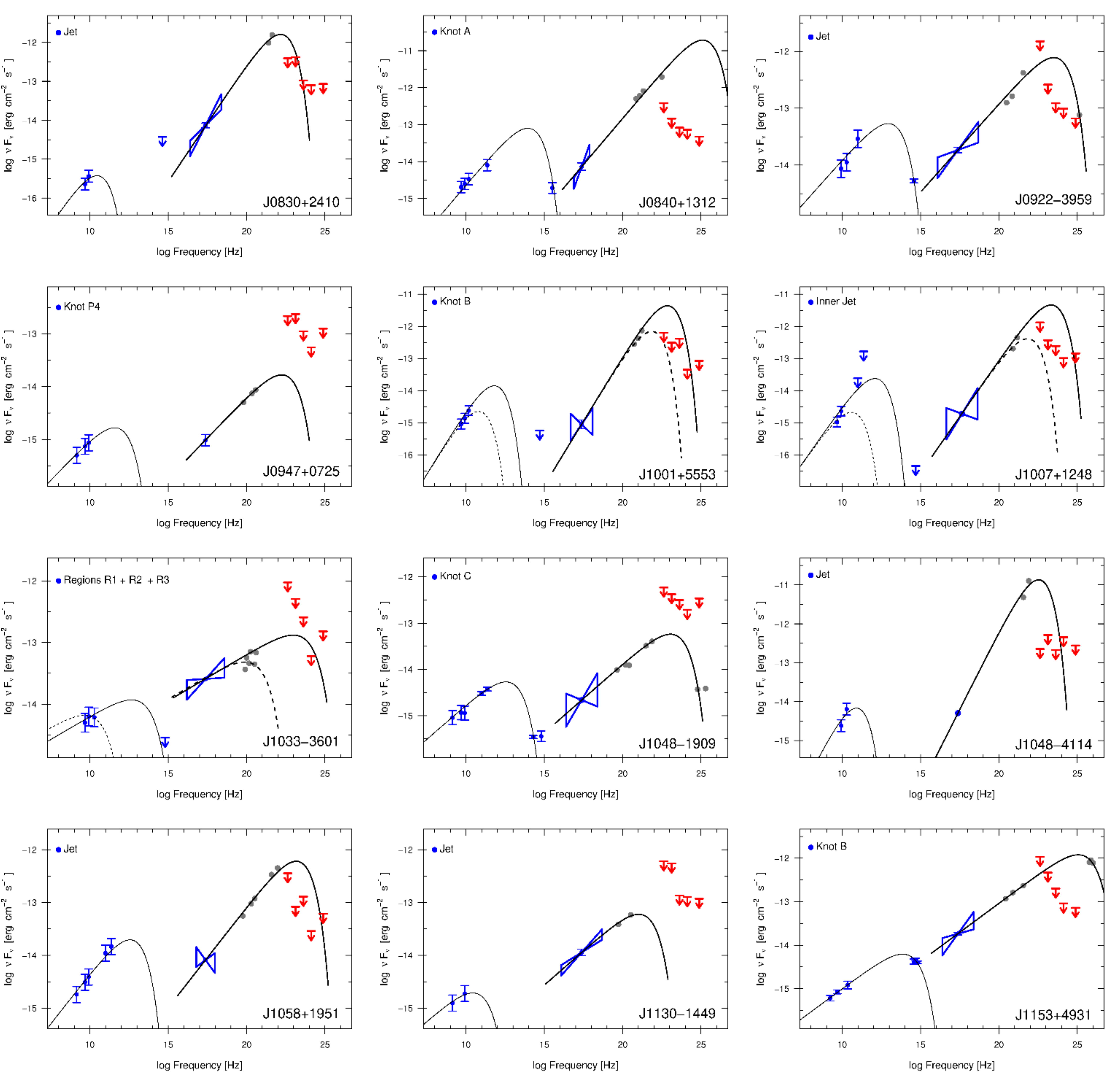}
	\caption{This is a continuation of Figure~\ref{fig:sed1}.}
	\label{fig:sed2}
\end{figure*}

\begin{figure*}
	\includegraphics[scale=0.33]{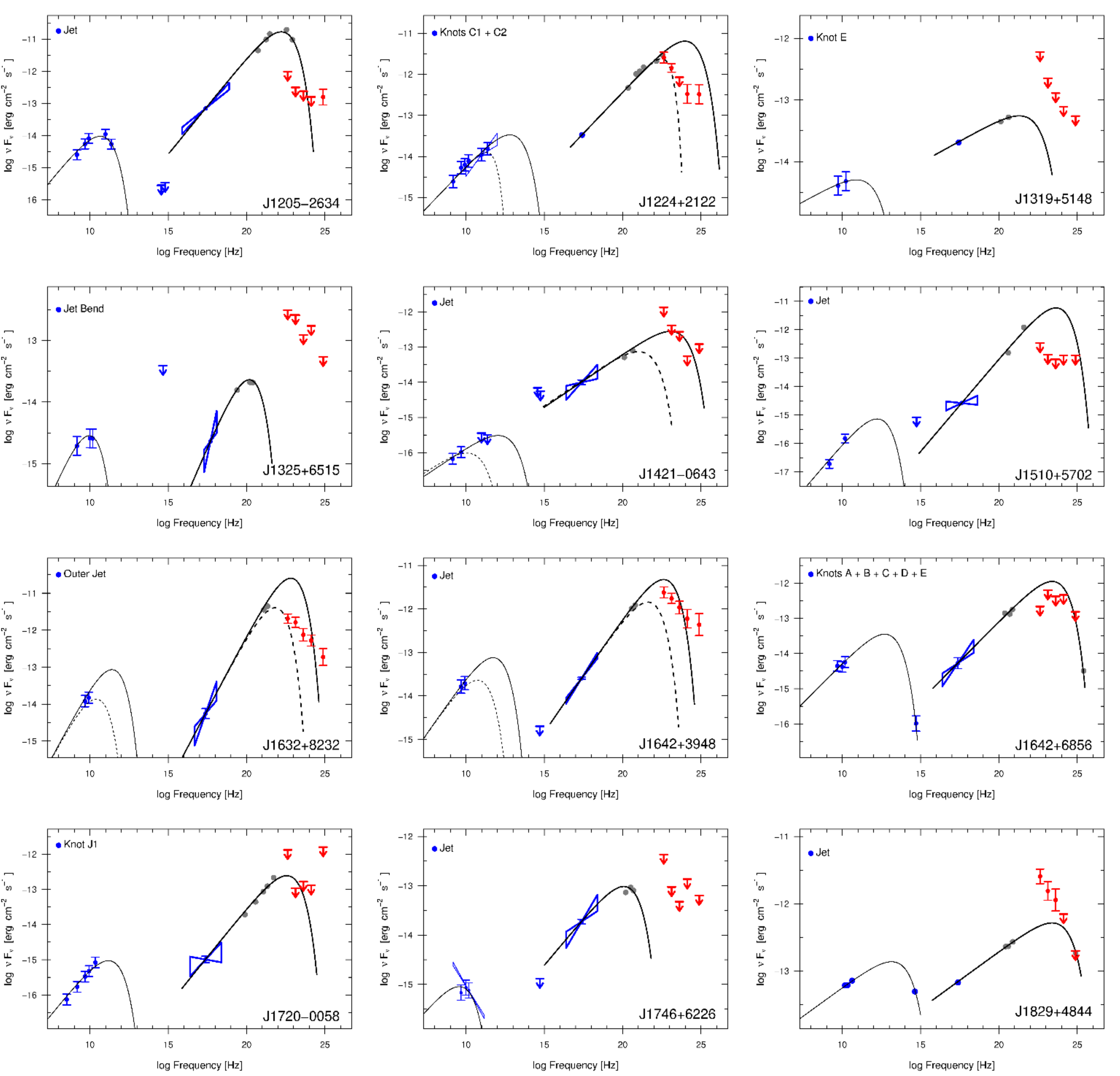}
	\caption{This is a continuation of Figure~\ref{fig:sed1}.}
	\label{fig:sed3}
\end{figure*}

\begin{figure*}
	\includegraphics[scale=0.33]{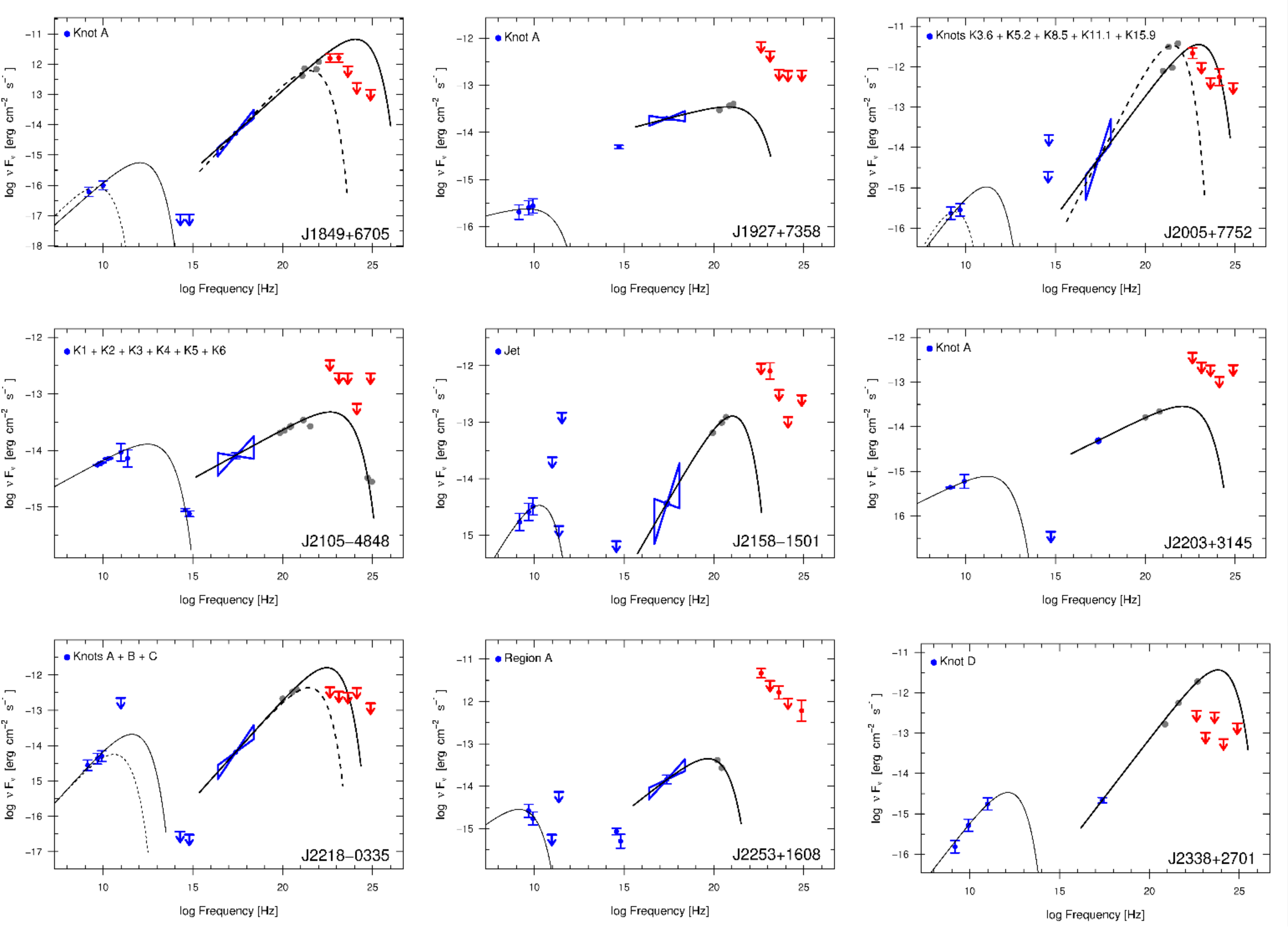}
	\caption{This is a continuation of Figure~\ref{fig:sed1}.}
	\label{fig:sed4}
\end{figure*}


\bsp	
\label{lastpage}
\end{document}